# On the Origin of Otho-Gardenhose Heliospheric Flux

Mike Lockwood* (1), Mathew J. Owens (1), and Allan Macneil (1)

*(1) Department of Meteorology, University of Reading UK;*

**Abstract**. Parker spiral theory predicts that the heliospheric magnetic field (HMF) will have components of opposite polarity radially toward the Sun and tangentially antiparallel to the solar rotation direction (i.e. in, Geocentric Solar Ecliptic (GSE) coordinates, with $B_X/B_Y < 0$). This theory explains the average orientation of the HMF very well indeed but does not predict the so-called "ortho-gardenhose" (hereafter OGH) flux with $B_X/B_Y > 0$ that is frequently observed. We here study the occurrence and structure of OGH flux, as seen in near-Earth space (heliocentric distance $r = 1$ AU) by the Wind and ACE spacecraft (for 1995-2017, inclusive) and by the Helios-1 and -2 spacecraft at $0.29$ AU $< r ≤ 1$ AU (for December 1974 to August 1981), in order to evaluate the contributions to OGH flux generation of the various mechanisms and factors that are not accounted for in Parker spiral theory. We study the loss of OGH flux with increasing averaging timescale $\tau$ between 16 seconds and 100 hours and so determine its spectrum of spatial/temporal scale sizes. OGH flux at Earth at sunspot minimum is shown to be more common than at sunspot maximum and caused by smaller-scale structure in the HMF (with a mode temporal scale at a fixed point of $\tau_{mp} \approx$ 10hrs compared to $\tau_{mp} \approx$ 40hrs for sunspot maximum, corresponding to about 5.5° and 22° (respectively) of heliocentric angular width for co-rotational motion or $21R_\odot$ and $84R_\odot$ for radial solar wind flow (where $R_\odot$ is a mean solar radius). OGH generated by rotating the HMF through the radial direction is also shown to differ in its spectrum of scale sizes from that for OGH generated by rotating the HMF through the tangential direction – the former does not contribute to the "excess" open heliospheric flux at a given $r$ but the latter does. We show that roughly half of the HMF deflection from the ideal Parker spiral needed to give the observed occurrence of OGH at Earth occurs at $r$ below 0.3 AU. By comparing the Helios and near-Earth data we highlight some questions which can be addressed by the Parker Solar Probe mission which will study the HMF down to $r = 0.046$ AU. We suggest that with decreasing heliocentric distance, Probe will detect decreased OGH field due to draping around transient ejecta, such as blobs and coronal mass ejections, but increasing structure in the radial field within traditional HMF sectors that are remnant Alfvénic disturbances in outflow regions from coronal reconnection sites.

1. Introduction

The Archimedean spiral in the Heliospheric Magnetic Field (HMF), known as the Parker spiral, was first proposed by *Parker* (1958). The theory is based on consideration of the effect of magnetic flux that is frozen-in in radial solar wind flow and dragged out of the solar atmosphere (*Ness and Wilcox*, 1964) whilst being rooted in the rotating solar corona. The theory is very successful predicting the average orientation of the field that is observed at a wide variety of locations in the heliosphere (*Behannon*, 1978, *Burlaga et al.,* 1982; *Bruno and Bavassano*, 1997; *Forsyth et al.*, 2002; *Jackman et al.,* 2008; *Borovsky*, 2010; *Owens and Forsyth*, 2013, *James et al.*, 2017). It predicts that the field will make an angle θ (called the "gardenhose angle") between the sunward radial and that θ will be between 0 and 90º for "T" flux (toward the Sun) and between −180º and −90º for "A" flux (away from the Sun); the value of θ depending on the local solar wind radial velocity $V$ and the heliocentric distance $r$. Note that section 1.1 shows that for the radial-field limits (θ = 0 for T field and θ = 180º, which is the same orientation as θ = −180º, for A field) the theory requires an infinite $V$ and for the tangential field limits (θ = 90º and θ = −90º) the theory requires $V$ = 0: hence these limits are only approached asymptotically in Parker spiral theory. At Earth ($r$ = 1AU, where AU is an Astronomical Unit), the average value of θ is predicted to be close to 45° or −135°, i.e. at the centers of the two allowed quadrants (see reviews by *Gazis*, 1996; *Borovsky*, 2010; *Owens and Forsyth*, 2013). However, as discussed in section 1.2, factors outside those considered by the theory cause some heliospheric field to lie outside the two allowed quadrants predicted by the theory and such flux is called "ortho-gardenhose" (OGH - as opposed to the field for which θ is within one of the two allowed quadrants which is called "gardenhose" or GH field). This paper studies the variation and structure of OGH field with $r$ and its spatial and temporal scales, with a view to defining its origins.

1.1. Parker spiral theory and the HMF gardenhose angle

The gardenhose angle that the heliospheric field makes with the direction radially toward the Sun is

$$\theta = \tan^{-1}(-B_Y/B_X) \quad (1)$$

where $B_X$ and $B_Y$ are the IMF components in the X (sunward) and Y directions of the Geocentric Solar Ecliptic (GSE) reference frame. We define GH flux as having

$$-\infty \leq B_Y/B_X \leq 0 \quad (2)$$

and OGH flux as having

$$0 < B_Y/B_X < \infty \quad (3)$$

These definitions mean that purely tangential field with no radial component ($B_X = 0$), predicted by Parker spiral theory in the limit of zero radial solar wind speed $V$, would be counted as GH (rather than OGH) orientation, as would purely radial T or A flux ($B_Y = 0$), predicted for infinite $V$. Values of $\theta$ in the first quadrant (Q1, $0 \leq \theta \leq 90°$) is T HMF ($B_X > 0$) in the GH orientation, $\theta$ in the second quadrant (Q2, $90 < \theta < 180°$) is A HMF ($B_X < 0$) with the OGH orientation, $\theta$ in the third quadrant (Q3, $-180° \leq \theta < -90°$) is A HMF with the GH orientation and $\theta$ in the fourth quadrant (Q4, $-90° < \theta < 0$) is T HMF with the OGH orientation.

Parker spiral theory gives a predicted gardenhose angle of:

$$\theta_p = \tan^{-1}(\omega r/V) + \theta_o \quad (4)$$

where $\omega$ is the angular rotation velocity of the corona and heliosphere and $\theta_o$ is zero for T HMF and $-180°$ for A HMF. Deviations from Parker spiral theory also occur in the form of latitudinal deflections of the field, but we here place no constrains on the out of ecliptic (Z) component or elevation angle of the field.

To demonstrate how good Parker spiral theory is in predicting the average HMF orientation in near-Earth space, Figure 1 presents polar histogram plots of the occurrence distribution of near-Earth heliospheric field (HMF) orientation in the ecliptic plane. The grey bars show the distributions of the observed field garden hose angle $\theta$ (from equation 1) in the GSE *X-Y* frame from 1-hour averages ($\tau = 1hr$) of the HMF for the years 1996-2017 (inclusive). Using the criterion for gaining an hourly mean of the IMF orientation that is accurate to within 5%, as derived by *Lockwood et al.*, (2018), this interval yields 161620 valid samples, an availability of 83.8%. The data are divided into 12 bins of equal sample numbers (13468 in each) between the percentiles of the distribution of the radial solar wind speeds, $V$ which are 315, 337, 355, 372, 390, 408, 428, 451, 483, 528 and 593 km s$^{-1}$ (see figure 2). The mauve histograms show the distributions of the orientation predicted from Parker spiral theory ($\theta_p$) for each of these intervals using equation (4) with the observed hourly mean radial solar wind speed $V$. The rotation of the means and modes of the distributions of both $\theta$ and $\theta_p$ towards radial with increasing $V$ matches that in the predicted field orientation. The OGH and GH sectors are shaded orange and green, respectively. OGH flux is most common for the lowest $V$ and less common for the highest $V$.

Figure 1 shows that the distributions in $\theta$ are continuous across the OGH/GH boundaries. Thus the processes that cause the spread in $\theta$ away from the predicted Parker spiral value, $\theta_p$, are the same as those that cause GH flux to become classified as OGH: the processes just need to be effective enough to make $|\theta-\theta_p|$ sufficiently large that one of the dividers between the four quadrants is crossed. Given the Parker spiral direction at a general $V$ and $r$ does not sit at exactly the centre of the GH quadrants, the proximity of the relevant quadrant boundary depends on which direction the field is deflected in. Figure 3 defines "Class A" and "Class B" OGH flux by the sense of rotation from the predicted Parker spiral (gardenhose angle $\theta_p$) to the local HMF direction (gardenhose angle $\theta$): Class A requires rotation through the radial direction, Class B rotation through the tangential direction. As shown in Figure 3, Class A rotation (pink arrow) is anticlockwise around the $Z_{GSE}$ axis when viewed from the northward side of the ecliptic; Class B rotation is in the opposite sense (pale blue arrow). The schematic also shows in the lower panels the corresponding Parker spiral field line and deflected field line in a wider-scale view looking down from above the north pole of the Sun.

Figure 2a gives the probability density function of the observed radial solar wind speed $V$ that yields the percentiles adopted in Figure 1, those percentiles being given by the vertical grey lines. The solid points in Figure 2(b) show the fraction of time that the HMF is in a OGH orientation ($T_{OGH}$, in black) for each of the 12 bins between these percentiles. This is also subdivided into Class-A and Class-B OGH flux by assuming the rotation sense gives the smaller angle needed to give the observed orientation (i.e., $|\theta-\theta_p| < 180°$). Given that Figure 1 shows that the Parker spiral orientation is closer to the tangential at low $V$ and closer to the radial at high V, it is not surprising that Class-B OGH flux is more common (blue points, $T_{OGH,B}$) at low $V$ but Class-A OGH flux becomes the more common at high $V$ (red points, $T_{OGH,A}$). OGH flux in general becomes less common at high $V$, as noted from Figure 1. The fraction of the total radial magnetic flux in the total, Class-A and Class-B OGH are presented in Figure 2c and will be discussed further in section (2). Figure 2d shows the magnitude of the average $B_X$, $B_Y$ and in-ecliptic HMF $(B_{XY} = B_X{}^2 + B_Y{}^2)^{1/2}$ components for the 12 averaging bins of $V$. Parker spiral theory predicts that as the spiral unwinds with higher $V$, the magnitude of the HMF will fall for a given strength of the source field at the top of the corona. Figure 3d shows that this does not occur and that $B_{XY}$ actually rises with increasing $V$. This is not necessarily a failing in Parker spiral theory and implies that the source field is stronger in regions of the corona that give faster wind. However, higher $V$ also implies greater variability in $V$ and a factor in the rise in $B_{XY}$ with average $V$ will be the compression of the field in Corotating Interaction Regions ahead of fast streams.

## 1.2. Sources of ortho-gardenhose HMF

Figure 4 presents schematics of some of the mechanisms that could generate OGH flux, given it is not predicted by Parker spiral theory. Figure 4(a) is a general schematic noting that waves, shocks and turbulence can all deflect the field over a range of temporal and spatial scales (*Burlaga et al.*, 1982; *Roberts et al.,* 1990; *Smith and Phillips*, 1996; *Horbury and Balogh*, 2001, *Ragot*, 2006; *Bruno and Carbone*, 2013, *Horbury et al.*, 2018). Figure 4(b) would arise if an emerging loop of magnetic flux leaving the solar corona spans a fast solar wind steam.  This could arise from footpoint exchange reconnections that cause such a configuration over an existing fast steam or from flow speed enhancement forming underneath a pre-existing coronal loop. Either causes the field in the stream to be dragged out further than the parts of the loop embedded in slower flow: this sort of effect was invoked as an explanation near-radial HMF the tail end of fast stream intervals by *Gosling and Skoug* (2002), *Jones et al.* (1998) and *Riley and Gosling* (2007). To the leading side of the stream in the schematic, the field tends to radial but as the fast stream is likely to be radial, it will probably remain in the GH quadrant. On the other hand, to the trailing side of the stream, the field could become OGH in nature (in the area shaded blue and this would be a Class-B rotation).  This means that the seeds of the deviation from Parker spiral orientation are sowed close to the sun in the solar atmosphere or even photosphere, as has been proposed for the magnetic "braiding" concept proposed by *Borovsky* (2008).  Another class of cause of deviations from Parker spiral orientation is the effects of draping pre-existing HMF over coronal ejecta released underneath it, as illustrated in Figure 4(c) (*Gosling and McComas* (1987); *Burlaga and Ness*, 1993; *Richardson and Cane*, 1996; *Smith and Phillips*, 1996; 1997). This could range from large CMEs (*McComas et al.,* 1998; *Kaymaz & Siscoe*, 2006) to smaller events down to small-scale transient blobs (*Sheeley et al.,* 1997; *Kilpua et al.,*

2009; *Rouillard et al.*, 2010a; 2010b; *Viall and Vourlidas,* 2015; *Kepko et al*., 2016). In this case the event is likely to expand as it propagates which means that the draped field to the westward side of the event may well be deflected past the radial direction and become Class-A OGH flux (the pink-shaded area). To the east of the event the draping is likely to give Class-B OGH flux. Figure 4(d) points out that both classes of OGH are likely to be formed in the outflow regions of reconnection sites. These may be relatively local sites in the heliosphere (as observed by *Phan et al.*, (2006) and *Mistry et al.* (2017)) or they may be back in the solar corona giving disconnected or folded flux (as inferred by *Owens et al.,* (2017; 2018)). Similarly, the leading edge of an erupting loop would give both class of OGH flux as it passed over a given location (Figure 4e) and that loop may have a flux rope form (e.g., *Chen et. al.*, 1997) which would allow the OGH flux regions to cover a more extensive region (Figure 4f). The key point we wish to make here is that this wide variety of processes can give OGH flux over a wide range of temporal and spatial scales.

### 1.3. The variation with heliocentric distance $r$

The launch of the Parker Solar Probe (*Fox et al*., 2106) gives us an opportunity to study the variation of OGH flux, and the gardenhose angle distribution in general, closer into the Sun and so understand more about how and where deviations from the Parker spiral are generated. This mission will study the HMF at exceptionally high time resolution and uniquely close to the Sun. The first three orbits have perihelion at $r = 0.16$ AU and later in the mission this a gradually reduced to $r = 0.046$ AU. In anticipation of these observations we here study data from the two Helios spacecraft that made measurements down to $r = 0.29$ AU.

### 1.4. Ortho-garden hose flux, folded flux and excess flux

*Owens et al.* (2008) surveyed radial HMF measurements throughout the heliosphere and found that the modulus of the radial component $|B_r|$ was largely independent of heliographic latitude (as expected from consideration of tangential magnetic pressure in the low-β solar wind close in to the Sun (*Suess and Smith*,1996)). However, these authors found a consistent rise in $|B_r|$ with radial distance. *Lockwood et al.* (2009a; 2009b) deployed the term "excess flux" for the difference between the total unsigned open solar flux that leaves the top of the solar atmosphere (the "coronal source surface" by convention usually taken to be at $r = 2.5 R_\odot$), $[4\pi r^2 |B_r|]_{r=2.5R_\odot}$, and the total unsigned flux threading a sphere at a general heliocentric distance $r$ in the heliosphere, $[4\pi r^2 |B_r|]_r$ and noted the results of *Owens et al.*

means that this "excess flux" increases with $r$. *Smith* (2011) argued this was merely an artefact of taking the modulus of the radial field and that excess flux violated Maxwell's equations by requiring magnetic monopoles between $2.5R_\odot$ and $r$. This is not the case because a heliospheric field line of a given polarity can fold back on itself (so called "folded flux") such that it threads the surface at $r$ a total of $n$ times, where $n$ is 3, or indeed any larger odd number, whereas it threads the surface at $2.5R_\odot$ just the once. Smith attributed the excess flux to an unspecified "noise" that grew with $r$ and so advocated avoiding the use of the modulus by averaging over T and A sectors in the field. However, the problem with this method was pointed out by *Lockwood and Owens* (2013), namely that there was a large uncertainty in deciding what was a true sector boundary (i.e. a T/A field polarity reversal at a given observation point that is known to map all the way back to the coronal source surface) and so this could not be done routinely. (In fact, that can be done in the presence of detectable unidirectional strahl electron flows because, although the field reverses at a true sector boundary the electron flow direction is always away from the Sun, whereas for folded flux both field and the strahl electron flow reverse direction – see below). *Lockwood et al.* (2009a) used the third perihelion pass of the Ulysses spacecraft to show that the excess flux mainly arose in the streamer belt where variability in solar wind velocity is high, and *Lockwood et al.* (2009b) developed a kinematic correction to allow for this effect which *Lockwood and Owens* (2009) showed explained the data survey as a function of $r$ by *Owens et al.* (2008) very well.

More recently, *Owens et al.* (2017) have been able to use strahl electron data to identify folded flux and so subtract its contribution to the open solar flux. The results are consistent with the kinematic correction of *Lockwood et al.* (2009b) to within the computed uncertainties; however, it is also true that the kinematic correction consistently gives slightly larger excess flux values than the strahl/folded flux method.

The point we wish to make here is that folded flux may often be associated with Class B OGH flux. Class A OGH flux makes no contribution to excess flux because by rotating through the radial direction the field line still only threads the spherical sphere of radius $r$ the one time. But Class B OGH flux does because it generates field lines that thread the surface multiple times and one in three of those crossings are in an OGH orientation. However, note that not all folded flux gives Class B OGH flux: if the field line folding is by an angle greater than $180°−θp$, then the flux is folded back on itself and so changes its T/A polarity but remains in a GH orientation: hence this would contribute to excess flux but not to Class B OGH. Therefore, although Class B OGH always contributes to excess flux, not all excess flux involves Class B OGH flux.

2. Analysis

For data on near-Earth space ($r$ = 1AU) we here employ the Omni composite of data at 1-minute resolution (made available by the Space Physics Data Facility (SPDF), NASA/Goddard Space Flight Center from https://omniweb.gsfc.nasa.gov/ow_min.html) (*King and Papitashvili*, 2005). We use data for 1995-2017 when the ACE and WIND spacecraft provide near continuous observations of the HMF.  This yields a total of 11046240 valid HMF samples which is an availability of 97.87% so the data are indeed very nearly continuous. We extend some studies in resolution down to 16s using 39446983 16-second samples from the ACE magnetometer (*Smith et al*, 1998) taken between 1998 and September 2018 (an availability of 99.86%). These data are made available by SPDF via CDAWeb (https://cdaweb.sci.gsfc.nasa.gov/index.html).

We also employ data from the Helios 1 and 2 spacecraft which executed orbits from $r$ = 0.29 to 1 AU with roughly a six-month orbital period.  We employ solar wind flow data from the plasma experiments (*Rosenbauer et al.,* 1977) and HMF observations from the magnetometers (*Neubauer et al*., 1977) on board both craft. The datasets have (approximately) 40-second resolution, with some datagaps being caused by loss of telemetry when the spacecraft were behind the Sun. The data were also downloaded from CDAWeb (from the same URL as given above) and were linearly interpolated onto regular times one minute apart and any 1-minute point that was removed from a valid data point by more than 40 seconds was treated as a data gap. This yielded 1517731 1-minute HMF samples from Helios 1 (which made 13 orbits between December 1974 and June 1981: an average data availability of 44.1%) and 895185 1-minute HMF samples from Helios 2 (which made 7 orbits between January 1976 and May 1979: an average data availability of 51.1%).  The Helios data commence shortly before the sunspot minimum between solar cycles 20 and 21 and end at the peak of cycle 21: hence they cover a sunspot minimum and a rising phase of the solar cycle. The daily sunspot number (we use the new version of the international sunspot number) vary between zero and 428 in this interval; by way of comparison, the largest value in the record, which starts in 1818, is 528 on 26[th] August 1870. The mode and mean values of the distribution for the times of all Helios samples (from both craft) are zero and 112.2, respectively. Hence the Helios data are dominated by sunspot minimum conditions, but do contain some samples at all activity levels up to sunspot maximum.

## 2.1. Observations of the Parker spiral and OGH flux at $r$ = 1 AU

In this section we continue the investigation of 1-minute Omni samples of the HMF in near-Earth space that was used to introduce the Parker spiral in section 1.1 of the present paper.

The black lines in Figure 5 shows distributions of θ for the near-continuous IMF data available for 1996 to 2017 (inclusive), these 22 years giving 11046240 valid one-minute averages ($\tau$ = 1 min., top panel) and 6428 30-hour averages ($\tau$ = 30 hr., bottom panel). In both cases, the number of valid samples in 4°-wide bins of gardenhose angle θ are counted, $N$, and normalised to the sum for all 90 bins, $\Sigma N$. The red lines are for three years around the sunspot maxima (the years 2000, 2001, 2002, 2012, 2013 and 2014, which yield 2891693 1-min samples and 1753 30-hr samples) and the blue lines are for 3 years around sunspot minimum (the years 1996, 1997, 2007, 2008, 2009 and 2017 which yield 2424326 1-min samples but only 1751 30-hr samples: two 30-hr intervals are lost because these sunspot minimum data come from three intervals, not the two used for sunspot maximum). The four quadrants are marked in the upper panel and clear peaks are seen in the GH quadrants (Q1 and Q3) in both cases. Note that noise is greater in (b) than in (a) because of the fewer samples. The averaging timescale $\tau$ has a very clear effect on the occurrence of OGH flux (quadrants Q1 and Q3) as at $\tau$ = 30hrs it is very rare, whereas at $\tau$ = 1 min even its minimum occurrence is about a quarter of the maxima in the GH quadrants. Comparison of the red and blue histograms in (a) shows that OGH flux (in Q2 and Q4) is, surprisingly, more common at sunspot minimum than at sunspot maximum (and so the peaks in Q1 and Q2 are lower because the histograms are normalised).

The GH peaks for T flux (Q1) and A flux (Q3) are not symmetrical, with the A peak being larger than the T one for sunspot maximum and the overall average. This is despite the data covering a whole Hale cycle (22 years) and hence both polarities of solar polar fields with equal weighting in terms of time. However, the two solar maxima in this interval are not of equal amplitude in terms of solar activity (solar cycle 24 being considerably weaker in sunspot number than cycle 23) and this may have introduced an asymmetry in the case of sunspot maximum. For sunspot minimum, the asymmetry is the other way round with the T peak (Q1) being larger than the A peak (Q3) – this may also be due to an asymmetry in solar activity for the two polar field polarities with the minimum between cycles 23 and 24 being considerably deeper and longer-lived than that between cycles 22 and 23. This asymmetry is the same in the 1-minute and 30-hour data. In this paper we are concerned with the effect of averaging timescale $\tau$, in particular on OGH flux, and discussion of asymmetries in the GH flux shown in Figure 5 will be left to a later paper.

Increasing the averaging timescale $\tau$ removes OGH flux because within a period of larger $\tau$, intervals of the non-GH polarity $B_X$ or $B_Y$ (giving positive-polarity HMF $B_X$/$B_Y$) that are seen at low $\tau$ are cancelled out. Thus studying the variation of the occurrence of OGH with $\tau$ reveals the spectrum of temporal scales at which the OGH exists at a given point. This could reflect stable longitudinal spatial structure in the heliosphere that is moved over the spacecraft with the Carrington rotation of the heliosphere or transient radial structure that is propagated over the spacecraft by the solar wind flow. In general, both will contribute and so we refer to the study of OGH as a function of $\tau$ as revealing its spatial and temporal structure. We can study the occurrence frequency of OGH orientations in this way, but often of greater interest is the amount of magnetic flux contained in the OGH sectors, and for studies of folded flux and excess flux, it is the flux in the radial direction that is most important. In section 2.2 we define how this is quantified.

## 2.2. Helios observations of the Parker spiral at $r < 1$ AU

Figure 6a presents the distributions of $\theta$ seen by the Helios 1 and 2 craft in 8 non-overlapping ranges of heliocentric distance, $r$, that are 0.1 AU wide and centered on $r$ of [0.3:0.1:1.0] AU. Histograms of the normalised number of samples in bins of gardenhose angle $\theta$ that are 4° wide, $N/\Sigma N$, are colour-coded by the range of $r$ using the key given. The evolution of the Parker spiral angle with $r$, expected from equation (4) can clearly be seen, with the distribution closest to the Sun ($r < 0.35$AU, in black) peaking closest to the radial direction, and the average gardenhose angle increasing with increasing $r$. The width of the distributions also increases a little with $r$. This can be seen most clearly in (b) which plots the distributions of the deviations from the predicted Parker spiral angle $\delta\theta = \theta - \theta_p$ for the same ranges of $r$ and using the same colour scheme. Given that $\theta_p$ is close to 45° for near Earth space, the fractions of the distributions with $|\delta\theta| \geq 45°$ is of interest because this is roughly sufficient deviation to cause OGH at $r = 1$ AU. This fraction rises linearly from 0.16 at $r = 0.3$ AU to 0.32 at $r = 1$ AU. On this basis we can say that roughly half of the at HMF deflection needed to give the observed occurrence of OGH $r = 1$ AU occurs at $r < 0.3$ AU and half occurs between $r = 0.3$ AU and 1 AU.

Figure 6c gives the variations with $r$ of the means and standard deviations of the $\delta\theta$ distributions shown in Figure 6b (respectively $<\delta\theta>$ and $\sigma_{\delta\theta}$), the black lines being the best least-squares linear regression in each case. The standard deviations $\sigma_{\delta\theta}$ confirm the broadening described above. The average values $<\delta\theta>$ are less consistent but show a marked tendency to be negative. This means that the Helios craft were detecting a tendency towards "underwound" HMF (with $\theta < \theta_p$). This has been noted by several authors particularly in Ulysses data at $r$ around 5 AU and within corotating rarefaction regions where deviations from the predicted Parker spiral angle can be as large as 30° (*Murphy et al.,* 2002). *Schwadron* (2002) modelled the underwound HMF resulting from magnetic footpoint motion at the Sun. The resulting field strays further from the Parker spiral angle with radial distance (see their Figure 3). This explains the relatively small deviations observed here by Helios, compared to the large 30 degree deviations observed by Ulysses. Note that sub-Parker spiral/underwound HMF can never contribute to OGH flux is because it is always deflected towards the radial, and can never cross out of a GH sector.

Using the variation of $<\delta\theta>$ and $\sigma_{\delta\theta}$ and extrapolating towards the Sun using the linear regressions, shown in Figure 6d, along with the predicted variation of $\theta_p$ with $r$, we can predict what the gardenhose angle distribution might look like at the smaller $r$ that the Parker Solar Probe will access. The results are shown in Figure 7. In order to predict $\theta_p$ as a function of $r$ using equation (4), we employ the model radial solar velocity profile $V(r)$ shown in Figure 7c, which is a 5th–order polynomial fit to the observed average velocities in the Helios data (solid points) and values at $r$ of $6R_\odot$ and $26R_\odot$ (open circles) derived by applying Fourier motion filters to SOHO/LASCO C3 movies observed from 1999 to 2010 by *Cho et al.* (2018). Figure 7a colour contours the derived p.d.f. as a function of garden hose angle $\theta$ with heliocentric distance $r$ using Gaussian distributions of the deviations from the Parker spiral direction of mean $<\delta\theta>$ and standard deviation $\sigma_{\delta\theta}$. The total toward and away distributions are assumed to be symmetrical. Figure 7b shows the predicted fraction of time that the field is in an OGH orientation, $T_{OGH}$: the black line is for all field, the red line for Class A OGH field and the blue line for Class B OGH field (see Figure 3). The areas shaded grey, pink and pale blue around the black, red and blue lines are the uncertainties introduced by the uncertainty in $<\delta\theta>$. As would be expected, the fraction of Class B OGH falls and of Class A OGH rises with decreasing $r$ as $\theta_p$ tends to zero.

We note that the assumption that the variation of $\sigma_{\delta\theta}$ remains linear is a major one that may well not be valid as, for example, turbulent perturbations of the plasma and field have been observed to grow rapidly between around $r$ of $15R_\odot$ and $69R_\odot$ (0.7-0.32AU) by *DeForest et al.* (2016) using images from the HI instruments of the STEREO spacecraft. Parker Solar Probe will reach down to $r$ of 0.046 AU, and so will study this region in-situ for the first time. Deviations of the pattern in Figure 7a from that for Solar Probe Data (or a lack of them) will help us define the processes giving us the spread in $\theta$ that arises $r < 0.3$ AU. Of particular interest for the present authors will be the reconnection of open flux with coronal loops. *Owens et al.* (2018) have recently used isotopic abundances to infer that this can occur right down into the low corona and that the subsequent evolution of the field lines gives OGH flux (they propose a more complex variant of Figure 4d). Because we can see no obvious reason why this should not also occur for more distended loops or at greater $r$ in the corona or inner heliosphere, we predict that at the lowest $r$ Solar Probe is likely to find considerable mixtures of near-radial T and A field within a classic HMF sector. The outflow exhaust regions from reconnection sites will be Alfvénic structures and if reconnection outside the corona and in the heliosphere is a factor as $r$ increases, the probability of outward flow would increase and the probability of inflow (or very slow outflow) will decline. Over larger distances the strength of these Alfvénic structures will decline and the dominant field polarity of the sector should begins to emerge more strongly and this change would be accompanied by a rise in the Class A OGH flux and of the gardenhose angle distribution width, $\sigma_{\delta\theta}$.

Lastly, Figure 8 compares the distributions of $\theta$ at $r < 0.39$ (i.e. the innermost 0.1AU covered by the Helios mission, shown in panel b) with those for $r = 1$ AU (the Omni data, shown in panel a). In both cases the black lines are for 1-minute data and the red lines are for 30-hour averages. As for Figures 1, 2 and 5, the orange and green bars at the top of each panel mark the OGH and GH sectors, respectively. In both cases, the OGH flux has almost vanished for $\tau = 30$ hours. For both $\tau$, the distribution is broader for $r = 1$AU than for $r < 0.39$ AU. (This figure will be discussed again later in connection with Figure 13.)

## 2.3. Ortho-gardenhose Flux Fraction

In this section, we describe how we quantify the fraction of the total unsigned near-Earth heliospheric magnetic flux that is in an OGH orientation. We study this ratio of the total unsigned flux values because both the numerator and denominator of this ratio rise and fall over the solar cycle with the total open solar flux. This also means that the decrease in the total unsigned flux with $r$ (predicted by Parker spiral theory to be a $r^{-2}$ decrease but not, in general, valid because of Class B OGH departures from Parker spiral orientations and because of "folded" GH flux both give "excess flux" (*Lockwood et al.*, 2009a; 2009b; *Owens et al.,* 2018)) is automatically accounted for.

Consider a band of unit length in the GSE Z-direction that is around the Sun at heliocentric distance $r = r_1 = 1$ AU which therefore has a surface area $2\pi r_1$. The total flux sum of inward and outward magnetic flux (i.e. the unsigned flux) threading the surface of this band for data averaged over a timescale $\tau$ is

$$\int_{n\,yr} |<B_X>_\tau| \, dt = \int_{n\,yr} |<[B_X]_{GH}>_\tau| \, dt + \int_{n\,yr} |<[B_X]_{OGH}>_\tau| \, dt \qquad (5)$$

Where the integral is over a whole number ($n$) of years such that Earth and its L1 satellites have moved around the band an integer number of times. We use the years 1996-2017 (inclusive, i.e. $n = 22$) which equals 299.75 Carrington rotation periods (which is 300 to within an accuracy of 0.08%). Thus our choice of $n$ covers approximately one whole Hale solar cycle, 22 whole solar orbits of the L1 spacecraft and very close to 300 whole Carrington mean solar rotations. $[B_X]_{GH}$ is the (sunward) $B_X$ field component for field for which $\theta$ is in one of the two GH quadrants and $[B_X]_{OGH}$ is the $B_X$ field component field for which $\theta$ is in one of the two OGH quadrants. The use of the modulus means that both inward and outward flux are included but because $B_X$ can have either polarity (T and A flux) all terms depend on the averaging timescale, $\tau$. The fraction of the total radial flux that is in the OGH orientation is

$$F_{OGH}(\tau) = \int_{n\,yr} |<[B_X]_{OGH}>_\tau| \, dt \,/ \int_{n\,yr} |<B_X>_\tau| \, dt = <|<[B_X]_{OGH}>_\tau|> \,/ <|<B_X>_\tau|> \qquad (6)$$

We use the 1-minute Omni data of the component $B_X$ and determine $[B_X]_{OGH}$ (which equals $B_X$ if $\theta$ is in the OGH quadrants Q2 and Q4 but zero if $\theta$ is in the GH quadrants Q1 and Q3). These are then averaged into intervals $\tau$ to give $<[B_X]>_\tau$ and $<[B_X]_{OGH}>_\tau$. The modulus is then taken and the average of those modulus values for all the intervals of duration $\tau$ taken.

The ratio then yields $F_{OGH}$ which will depend on $\tau$ which is varied between 2 minutes and 100 hours in 1000 steps that are multiples of 1 minute and are spaced quasi-logarithmically. This analysis was also extended back to sub-minute timescale using the 16-second ACE data which is averaged over intervals of $\tau = 32$ and 48 seconds.

Figure 9 shows the effect of averaging on garden hose flux and on $F_{OGH}$. The black line in the top panel shows the unsigned radial flux fraction for OGH flux, $F_{OGH}$, as a function of $\tau$. The red and blue lines show the variations of the corresponding values for data taken in 3 year intervals around sunspot maximum and sunspot minimum, respectively. As expected the cancellation of opposite-polarity $B_X$ and $B_Y$ within averaging intervals causes a reduction in the width of the $\theta$ distribution peaks and a fall in the OGH flux with increasing $\tau$ (plotted in Figure 9a as $\log_{10}(\tau)$ to reveal the changes a low $\tau$). This fall is not the same at sunspot maximum as at sunspot minimum, and as noted above, there is more OGH flux at sunspot maximum at low $\tau$. (Note that Figure 9a shows that at $\tau$ above about 8 hours there is actually more OGH flux at sunspot maximum, this is explained below). Figure 9b shows $N/\Sigma N$ colour-contoured as a function of $\theta$ and $\log_{10}(\tau)$, where $N$ is the number of samples in 90 bins of $\theta$ that are $d\theta = 4°$ wide, and $\Sigma N$ is the sum of $N$ over all bins. This plot demonstrates how increased $\tau$ sharpens and raises the peaks in the distribution of GH flux as the OGH flux is averaged out.

## 2.4. Spectra of OGH flux scale sizes

Figure 9(a) demonstrates that increasing the averaging timescale from $\tau_1$ to $\tau_2$ causes a decrease in the OGH unsigned radial flux fraction as OGH is cancelled. However the change in gradient in the plots tell us that some changes from $\tau_1$ to $\tau_2$ cause more flux loss than others, which in turn tells us about the amount of structure in the HMF giving OGH that is of scale between $\tau_1$ to $\tau_2$. This structure is cancelled by averaging in the time domain but as the time series is at a fixed point the temporal variation tells us about larger-scale temporal variations in the heliosphere around the point, or radial spatial structure that is moved over the location by the solar wind flow, or tangential spatial structure rotated over the point by Carrington rotation of the heliosphere. For the radial structure, the spatial scale is related to the temporal scale $\tau$ by $V\tau$ (where $V$ is the solar wind speed) and for the tangential structure it is related by $\omega r\tau$ (where $\omega$ is the angular velocity of heliospheric rotation relative to the observing platform).

Figure 10 analyses the spectrum of scales of all OGH flux (Class A and Class B flux combined) at $r = 1$AU. The plots on the left hand side use an x axis that is linear in averaging timescale $\tau$, the right hand plots are the same, other than the x axis is logarithmic. Figures 10a and 10b show $\Delta F$, the fraction of the total radial magnetic flux that is present at 16 s resolution that is lost by cancellation by averaging over an interval $\tau$ (by definition zero for the fundamental resolution of the data which is $\tau = 16$ s). The grey line is the observed value and the mauve line a 6th order polynomial fit to smooth out the small numerical noise associated with the start times of the averaging intervals relative to features in the data. This noise gets greater at large $\tau$ because the number of available samples falls. The polynomial fit is actually generated using the $\log_{10}(\tau)$ variation shown in part (b) and then plotted on a linear scale in (a) as this captures the variation at small $\tau$ more satisfactorily. Parts (c) and (d) show the flux lost in increasing the averaging time between $\tau_1$ and $\tau_2$, $\delta F$: in (c) the area shaded grey is made up of vertical bars of height $\delta F$ between $\tau_1$ and $\tau_2$ and in (d) the bars are again of height $\delta F$ but between $\log_{10}(\tau_1)$ and $\log_{10}(\tau_2)$. In this way, the total area shaded grey in both plots equals the total fraction of radial unsigned OGH flux data that is lost by averaging over intervals of duration $\tau = 100$ hrs and the shapes of the grey areas allows us to identify what timescales $\tau$ contribute most to this loss, and hence gives the spectrum of scale sizes of the OGH flux. The linear plot (c) allows us to see the structure at high $\tau$ most clearly, the logarithmic plot (d) reveals structure at low $\tau$. In each panel vertical blue lines are drawn for $\tau = 1$ hr and $\tau = 50$ hours for reference. To interpret the spectra we note that $\tau = 1$ hr. corresponds to a tangential structure scale length of $1.55\times10^6$ km (i.e., 0.01 AU, $2.2 R_\odot$, or 243 $R_E$, where $R_\odot$ is a solar radius and $R_E$ is a mean Earth radius or an angular width of 0.6° subtended at the Sun); $\tau = 50$ hr. corresponds to a tangential structure scale length of $7.74\times10^7$ km (i.e., 0.5 AU, $111 R_\odot$, 12145 $R_E$, or an angular width of 30°). The distances apply to radial structure for a radial solar wind velocity of 430 km s$^{-1}$ (which makes $\theta_p = 45°$ for T flux and $-135°$ for A flux).

Figures 10c and 10d show a clear peak at around $\tau = 8$ hours which corresponds to a distance of about 0.08 AU or an angular width of about 5°. Figure 11 is the equivalent of Figure 10c and 10d for, again for $r \approx 1$AU and for: (top panels) data taken within 1 year of sunspot minimum; and (bottom panels) data taken within one year of sunspot maximum. The spectra are considerably different. Figures 11a and 11b show that the peak at $\tau = 8$ hours is a sunspot minimum phenomenon but is not seen a sunspot maximum for which, instead, there is an almost flat plateau rising up to a mode value of about $\tau = 40$ hours. We conclude that there are significantly more large structures contributing to OGH flux at sunspot maximum and because they last up to almost 2 days these appear to be the effect of large CMEs.

Figure 12 shows a similar difference in the data $r \approx 1$AU when the Class A and Class B OGH flux are compared. An obvious difference between the two is that there is less Class B OGH flux to lose, but this is not surprise because for Class B the field is deflected towards and past the tangential direction, for which the $B_X$ component is zero, whereas Class A is deflected toward and past the radial direction for which $|B_X|$ is large. Comparing the shapes of the spectra the peak around $\tau = 8$ hours is again present but only for the Class A flux and Class B exhibits a plateau between 8hrs and 50 hours, suggesting a wide range of ejecta, from blobs to large CMEs are involved.

## 2.5. Spectra of OGH flux scale sizes at $r < 0.39$ AU

Figure 13 repeats Figure 12 for the Helios data taken near perihelion. In order to keep sample sizes high enough at large $\tau$, the Class A and Class B bin widths have been increased, as shown in Figure 8b. This increases the number of samples but does not have a strong influence on the spectral shape. The pink and pale blue areas in Figure 8a define the Class A and Class B OGH flux. The join between them is at the orthogonal to the average Parker spiral direction. The ranges of θ covered by the pink and pale blue areas in Figure 8b have been expanded in width to cover some of the GH sectors as well as the OGH sectors to allow for the fact that the distributions in the GH sectors are narrower: again the divider between them is the orthogonal to the average spiral direction but the width of the A and B bins has been increased until the total fraction of $\tau = 30$ hr samples in such bins is the same as for the A and B ranges in the $r = 1$ AU data shown in Figure 8a. This somewhat arbitrary choice is only relevant to Figure 13 and the change to the bin width is made so ensure that the number of samples available for analysis at large $\tau$ (30-100 hours) is comparable in the two cases. The required Class A and Class B sector width is 72º in Figure 8b, as opposed to the bins of near 45º width in Figure 8a.

Because the spacecraft is at smaller $r$, (where the solar wind speed is lower) all the spatial distances for a given τ are decreased pro-rata, but the angular tangential widths for the same τ. The peak of the spectrum for Class A for these near-Sun observations is around 3 hours (corresponding to an angular width of just under 2º). This implies that the structures giving Class A OGH expand from about 2º to about 5º in propagating from about 0.3AU to about 1AU. *Kilpua et al.* (2009) report measurements of structures which they argue are consistent with being the in-situ manifestation of the blobs observed remotely (at ~3-30$R_\odot$) by *Sheeley et al.* (1997) and have sizes which are reasonably consistent with the dominant scale sizes found here. *Kepko et al.* (2016) report blobs which last ~1.5 hours which is a bit smaller than the scale of OGH flux that we find. This may, however not be inconsistent as the OGH scale is the scale of field line draping over the blob and not of the blob itself.

On the other hand, Parts (c) and (d) of Figure 13 are not greatly dissimilar from the corresponding plots in Figure 12, although rather than the plateau there is a suggestion of two peaks at τ near of 1 hour and 35 hours.

## 3. Discussion and Conclusions

The survey presented here shows there is there is an approximately linear growth in the non-Parker-spiral component of the HMF with radial distance. The available data suggests significant non-Parker-spiral fields, if not actually present at the source surface, are established in the very early evolution of the solar wind. By $r \approx 0.3$AU roughly half the deflection needed to give the OGH at Earth is present. In this paper, we have projected the trends to $r < 0.3$AU, using simple empirical extrapolation. However, remote sensing observations (*DeForest et al.*, 2016) suggest that that the deflection by waves and turbulence on small spatial scales grows rapidly between 0.07 and 0.3AU and so that extrapolation may well not be valid. Solar Probe will help define the role of waves and turbulence by observing the radial evolution of the fraction of flus that is OGH-orientated.

Perhaps surprisingly, OGH flux at sunspot maximum is found to be rarer than at sunspot minimum. This may well be because OGH flux generation by draping is more effective when events are isolated. We have also shown that the spectra of OGH scale sizes are significantly different at sunspot minimum and maximum. The peak at sunspot minimum is at a scale of about 10 hours which, if it were due to the draping over transients is broadly consistent with the effect of draping over, and/or release of, transient blobs (*Kilpua et al.*, 2009; *Kepko et al.*, 2016). From the remote sensing data, the occurrence of blobs appears to be rather even across the solar cycle in the ecliptic plane [e.g., *Luhmann et al.,* 2013] which makes it surprising that the small scale (~10hr) OGH structure at sunspot minimum is so very much greater than at sunspot minimum if transient blobs are the only cause.

At sunspot maximum, structure at a scale of 10 hours is again observed, but with much lower amplitude than at sunspot minimum. The overall spectrum has a quite different shape in this case, with amplitude increasing weakly with increasing timescale up to a mode value of about 40 hours. This is greater than the mode of the distribution of durations of Coronal Mass Ejection CME events which is of order 20 hours (*Mitsakou and Moussas*, 2014), but as for the blobs, this is likely to represent the difference between the characteristic scale of the draping region and the scale of the structure that the field is draped around.

We have also found a similar difference between OGH flux that is Class A (rotated through the radial) and Class B (rotated through the tangential). Figures 4b and 4c suggest that CMEs, blobs and fast streams will predominantly generate Class B OGH flux and this is consistent with the idea that these are the predominant drivers of sunspot-maximum OGH.

The Class A OGH flux at small scales (10hrs and less) is particularly common at sunspot minimum, and at both $r$ = 1AU and at $r$ = 0.3 AU. As illustrated by figure 4c, Class A OGH can be generated by CMEs because they expand as they propagate. However, because this effect would increase in magnitude and scale with increased $r$, this does not appear to be a candidate driver for this Class A OGH flux. A much more likely candidate is suggested by our studies of sunward strahl electron flows, namely that this Class A OGH is caused by magnetic reconnection in the corona, as shown schematically in Figure 4d. This makes some specific suggestions that can be tested by Parker Solar Probe. The Class A OGH flux in reconnection outflow regions will be Alfvénic disturbances emanating from the reconnection site and standing in the inflow regions. *Owens et al.* (2018) have recently used isotopic abundances to infer that this reconnection can occur right down in the low corona with open flux reconnecting with hot, dense coronal loops. Hence the outflows detected by Owens at al. are likely to be from the "interchange" reconnections that allow co-rotation of the corona. At greater $r$, at the coronal source surface and beyond, reconnections associated with the main (tilted) heliospheric current sheet (HCS) will be present and these give loss of open flux by disconnection at a rate that varies over the solar cycle with the tilt of the HCS (*Owens et al.*,

2011). *Lockwood et al.* (2017) point out that the HCS is highly unlikely to have a sharp inner edge at the source surface and propose that with decreasing $r$ within the corona, the main HCS increasingly breaks up into a network of smaller-scale sheets, an idea that can be merged with the "S-Web" concept of slow solar wind origin (*Antiochos et al.,* 2011). This would be a mixture of both disconnection and interchange reconnections and would cause a spreading the source locations of the reconnection outflows away from the sector boundaries and to within the HMF sectors. Hence we predict that at perihelion, Solar Probe is likely to find considerable mixtures of near-radial T and A field within the classic HMF sectors. These will usually be Alfvénic structures being the outflow regions of coronal reconnection sites and may often be associated with sunward strahl, OGH field orientations and, for the lowest $r$ interchange reconnections, hot and dense plasma. The reconnections, and their outflow regions will often be transient in nature but could be persistent, co-rotating structures such as on coronal hole boundaries. In general, these Alfvénic structures will be ironed out with increasing $r$ by the curvature force of the reconnected field lines; however, as pointed out by *Owens et al.* (2018), in the case of the lowest $r$ reconnections, the field rotation may be embedded in slow solar wind that is outrun by the fast flow on either side of it and the structure persists right out to, and past, 1AU. If for the majority of cases, the curvature force does iron-out the field structure, the dominant field polarity of the sector emerges more clearly as $r$ increases. If these interchange reconnections in the corona are indeed the origin of the Class A OGH detected here at both 1Ay and by the Helios spacecraft, Solar Probe would see the variations with $r$ predicted here by the simple model presented.

**Acknowledgements.** We are grateful to the instruments teams for the Helios plasma and magnetic field experiments, led by R. Schwenn and F. Neubauer and to by the Space Physics Data Facility (SPDF) at NASA/Goddard Space Flight Center for making these data available via CDAWeb (https://cdaweb.sci.gsfc.nasa.gov/index.html). We also than the teams of the corresponding instruments on the ACE satellite led by N.F. Ness and D. McComas and the data are again made available by SPDF (from the above URL). We also employ the Omni composite of data at 1-minute resolution (made available by SPDF from https://omniweb.gsfc.nasa.gov/ow_min.html). The work presented in this paper is supported by STFC consolidated grant number ST/R000921/1, the work of ML and MJO is also supported by the SWIGS NERC Directed Highlight Topic Grant number NE/P016928/1/

# References

Antiochos, S.K., J.A. Linker, R. Lionello, Z. Mikić, V. Titov, and T.H. Zurbuchen (2011) The Structure and Dynamics of the Corona—Heliosphere Connection, *Space Sci. Rev.*, **172** (1-4), 169-185, doi: 10.1007/s11214-011-9795-7

Behannon, K. W. (1978), Heliocentric distance dependence of the interplanetary magnetic field, *Rev. Geophys.,* **16** (1), 125–145, doi: 10.1029/RG016i001p00125.

Borovsky, J. E. (2010), On the variations of the solar wind magnetic field about the Parker spiral direction, *J. Geophys. Res.*, **115**, A09101, doi: 10.1029/2009JA015040.

Bruno, R. and Bavassano, B. (1997) On the winding of the IMF spiral for slow and fast wind within the inner heliosphere, *J. Geophys. Res.*, **24**, 2267–2270, doi: 10.1029/97gl02183

Bruno, R. and V. Carbone (2013) The Solar Wind as a Turbulence Laboratory, *Living Rev. Solar Phys.*, 10, 2. URL (accessed 1st August 2018): http://www.livingreviews.org/lrsp-2013-2, doi: 10.12942/lrsp-2013-2

Burlaga, L. F., and N. F. Ness (1993), Large-scale distant heliospheric magnetic field: Voyager 1 and 2 observations from 1986 through 1989, *J. Geophys. Res*., **98** (A10), 17451–17460, doi:10.1029/93JA01475.

Burlaga, L.F., Lepping, R.P., Behannon, K.W., Klein, L.W. and Neubauer, F.M. (1982) Large-scale variations of the interplanetary magnetic field: Voyager 1 and 2 observations

Chen, J., R.A. Howard, G.E. Brueckner, R. Santoro, J. Krall, S.E. Paswaters, O.C. St. Cyr, R. Schwenn, P. Lamy, and G.M. Simnett (1997) Evidence of an Erupting Magnetic Flux Rope: LASCO Coronal Mass Ejection of 1997 April 13. , *Astrophys. J. Lett*., **490**, L191 (40pp), doi: 10.1086/311029

Cho, I.-H., Y.-J. Moon, V.M. Nakariakov, S.-C. Bong, J.-Y. Lee, D, Song, H. Lee, and K.-S. Cho (2018) Two-Dimensional Solar Wind Speeds from 6 to 26 Solar Radii in Solar Cycle 24 by Using Fourier Filtering, *Phys. Rev.Lett.*, **121** (7), paper # 075101. doi: 10.1103/physrevlett.121.075101

DeForest, C.E., W.H. Matthaeus, N.M. Viall, and S.R. Cranmer (2016) Fading coronal structure and the onset of turbulence in the young solar wind, *Astrophys. J.*, **828**, 66 (16pp), doi: 10.3847/0004-637X/828/2/66

Forsyth, R.J., Balogh, A. and Smith, E.J. (2002) The underlying direction of the heliospheric magnetic field through the Ulysses first orbit, *J. Geophys. Res.*, **107** (A11), 1405, doi: 10.1029/2001JA005056

Fox N.J., M.C. Velli, S.D. Bale, R.Decker, A.Driesman, R.A. Howard, J.C. Kasper, J.Kinnison, M.Kusterer, D.Lario, M.K. Lockwood, D.J. McComas, N.E. Raouafi, and A. Szabo (2016) The Solar Probe Plus Mission: Humanity's First Visit to Our Star, *Space Sci. Rev.,* **204** (1–4), 7–48, doi: 10.1007/s11214-015-0211-6

Gazis, P. R. (1996) Solar cycle variation of the heliosphere. *Rev. Geophys.*, **34**, 379–402, doi: 10.1029/96rg00892

Gosling, J. T., and D.J. McComas (1987)  Field line draping about fast coronal mass ejecta: A source of strong out-of-the-ecliptic interplanetary magnetic fields, *Geophys. Res. Lett.*, **14** (4), 355 – 358, doi: 10.1029/GL014i004p00355

Gosling, J. T., and R. M. Skoug (2002) On the origin of radial magnetic fields in the heliosphere, *J. Geophys. Res.*, **107** (A10), 1327,  doi:10.1029/2002JA009434, 2002.

Horbury, T. S., and A. Balogh (2001), Evolution of magnetic field fluctuations in high-speed solar wind streams: Ulysses and Helios observations, *J. Geophys. Res.*, **106** (A8), 15929–15940, doi: 10.1029/2000JA000108.

Horbury, T.S., L Matteini, D. Stansby (2018) Short, large-amplitude speed enhancements in the near-Sunfast solar wind, *Mon. Not. of the Roy. Astron. Soc*., **478** (2), 1980–1986, doi: 10.1093/mnras/sty953

Jackman, C. M., R. J. Forsyth, and M. K. Dougherty (2008), The overall configuration of the interplanetary magnetic field upstream of Saturn as revealed by Cassini observations, *J. Geophys. Res.,* 113, A08114, doi: 10.1029/2008JA013083.

James, M.K., S.M. Imber, E.J. Bunce, T.K.Yeoman, M. Lockwood, M.J. Owens and J.A. Slavin (2017) Interplanetary magnetic field properties and variability near Mercury's orbit, *J. Geophys. Res. Space Physics*, **122**, 7907-7924, doi:10.1002/2017JA024435

Jones, G. H., A. Balogh, and R. J. Forsyth (1998), Radial heliospheric magnetic fields detected by Ulysses, *Geophys. Res. Lett.,* **25**, 3109– 3112, doi: 10.1029/98gl52259

Kaymaz, Z. & G. Siscoe (2006) Field-Line Draping Around ICMES, *Sol. Phys*., **239** (1/2), 437–448, doi: 10.1007/s11207-006-0308-x

Kepko, L., N. M. Viall, S. K. Antiochos, S. T. Lepri, J. C. Kasper, and M. Weberg (2016), Implications of L1 observations for slow solar wind formation by solar reconnection, *Geophys. Res. Lett.,* **43**, 4089–4097, doi:10.1002/2016GL068607.

Kilpua, E.K.J., J.G. Luhmann, J. Gosling, Y. Li, H. Elliott, C.T. Russell, L. Jian, A.B. Galvin, D. Larson, P. Schroeder. K. Simunac and G. Petrie (2009) Small Solar Wind Transients and Their Connection to the Large-Scale Coronal Structure, *Sol. Phys*., **256**, 327–344, doi: 10.1007/s11207-009-9366-1

King, J.H. and N.E. Papitashvili (2005) Solar wind spatial scales in and comparisons of hourly Wind and ACE plasma and magnetic field data, *J. Geophys. Res.,* **110**, A02104, 2005

Lockwood, M. and M. Owens (2009) The accuracy of using the Ulysses result of the spatial invariance of the radial heliospheric field to compute the open solar flux, *Astrophys. J*., **701** (2), 964-973, doi: 10.1088/0004-637X/701/2/964

Lockwood, M. and M.J. Owens (2013) Comment on "What causes the flux excess in the heliospheric magnetic field?" by E.J. Smith, *J. Geophys. Res*., **118** (5), 1880 1887, 2013. doi: 10.1002/jgra50223

Lockwood, M., M. Owens, and A.P. Rouillard (2009a) Excess Open Solar Magnetic Flux from Satellite Data: I. Analysis of the 3rd Perihelion Ulysses Pass, *J. Geophys. Res*., 114, A11103, doi:10.1029/2009JA014449

Lockwood, M., M. Owens, and A.P. Rouillard (2009b) Excess Open Solar Magnetic Flux from Satellite Data: II. A survey of kinematic effects, *J. Geophys. Res.,* **114**, A11104, doi: 10.1029/2009JA014450

Lockwood, M., M. J. Owens, S. M. Imber, M. K. James, E. J. Bunce, and T. K. Yeoman (2017), Coronal and heliospheric magnetic flux circulation and its relation to open solar flux evolution, *J. Geophys. Res. Space Physics*, **122**, 5870–5894, doi:10.1002/2016JA023644.

Lockwood, M., S. Bentley, M.J. Owens, L.A. Barnard, C.J. Scott, C.E. Watt, and O. Allanson (2018) The development of a space climatology: 1. Solar-wind magnetosphere coupling as a function of timescale and the effect of data gaps, *Space Weather*, in press, doi: 10.1029/2018SW001856 (2018)

Luhmann, J.G., G. Petrie and P. Riley (2013) Solar origins of solar wind properties during the cycle 23 solar minimum and rising phase of cycle 24, *J. Adv. Res*., **4**(3), 221–228, doi: 10.1016/j.jare.2012.08.008

McComas, D. J., J. T. Gosling, S. J. Bame, E. J. Smith, and H. V. Cane (1989), A test of magnetic field draping induced B z perturbations ahead of fast coronal mass ejecta, *J. Geophys. Res*., **94** (A2), 1465–1471, doi: 10.1029/JA094iA02p01465.

Mitsakou, E. and X., Moussas (2014) Statistical Study of ICMEs and Their Sheaths During Solar Cycle 23 (1996 – 2008), *Sol. Phys,* 289, 3137–3157, doi: 10.1007/s11207-014-0505-y

Mistry, R., J. P. Eastwood, T. D. Phan, and H. Hietala (2017) Statistical properties of solar wind reconnection exhausts, *J. Geophys. Res. Space Physics*, **122**, 5895–5909, doi:10.1002/2017JA024032.

Murphy, N., E. J. Smith, and N. A. Schwadron (2002) Strongly underwound magnetic fields in co-rotating rarefaction regions: Observations and Implications, *Geophys. Res. Lett.*, 29(22), 2066, doi:10.1029/2002GL015164.

Ness, N.F., and J.M. Wilcox (1964) Solar Origin of the Interplanetary Magnetic Field, *Phys. Rev. Lett.* 13, 461, doi: 10.1103/PhysRevLett.13.461

Neubauer, F.M., H. J. Beinroth, H. Barnstorf, and G. Dehmel (1977) Initial results from the Helios-1 search-coil magnetometer experiment, *Journal of Geophysics - Zeitschrift für Geophysik*, 42 (6), 599-614.

Owens, M.J. and R.J. Forsyth (2013) The Heliospheric Magnetic Field, *Living Rev. Solar Phys.*, **10**, 5. URL (accessed 1st November 2018), http://www.livingreviews.org/lrsp-2013-5, doi: 10.12942/lrsp-2013-5

Owens, M. J., Arge, C. N., Crooker, N. U., Schwadron, N. A. and Horbury, T. S. (2008) Estimating total heliospheric magnetic flux from single-point in situ measurements, J. Geophys. Res., 113 (A12). A12103, doi: 10.1029/2008JA013677

Owens, M.J., N.U. Crooker, and M. Lockwood (2011) How is open solar magnetic flux lost over the solar cycle?, *J. Geophys. Res.*, **116**, A04111, doi: 10.1029/2010JA016039

Owens, M.J., M. Lockwood, P. Riley and J. Linker (2017) Sunward strahl: A method to unambiguously determine open solar flux from in situ spacecraft measurements using suprathermal electron data, *J. Geophys. Res., Space Physics*, **122** (11), 10,980-10,989, doi: 10.1002/2017JA024631

Owens, M.J., M. Lockwood, L.A. Barnard, and A.R. MacNeil (2018) Generation of inverted heliospheric magnetic flux by coronal loop opening and slow solar wind release, *Astrophys. J. Lett.*, **868**, L14 (5pp), doi: 10.3847/2041-8213/aaee82

Parker, E.N., 1958, "Dynamics of the Interplanetary Gas and Magnetic Fields.", Astrophys. J., 128, 664–676. doi: 10.1086/146579

Phan, T.D., J.T. Gosling, M.S. Davis, R.M. Skoug, M. Øieroset, R.P. Lin, R.P. Lepping, D.J. McComas, C.W. Smith, H. Reme, and A. Balogh (2006) A magnetic reconnection X-line extending more than 390 Earth radii in the solar wind, *Nature*, 439(7073), 175-178, doi: 10.1038/nature04393

Ragot, B. R. (2006) Distributions of Magnetic Field Orientations in the Turbulent Solar Wind, Ap. J., 651 (2) 1209-1218, doi: 10.1086/507783

Richardson, I. G., and H. V. Cane (1996), Particle flows observed in ejecta during solar event onsets and their implication for the magnetic field topology, *J. Geophys. Res.*, **101** (A12), 27521–27532, doi:10.1029/96JA02643.

Riley, P., and J. T. Gosling (2007), On the origin of near-radial magnetic fields in the heliosphere: Numerical simulations, *J. Geophys. Res.*, **112**, A06115, doi: 10.1029/2006JA012210.

Roberts, D. A., M. L. Goldstein, and L. W. Klein (1990), The amplitudes of interplanetary fluctuations: Stream structure, heliocentric distance, and frequency dependence, *J. Geophys. Res.*, **95** (A4), 4203, doi:10.1029/JA095iA04p04203.

Rosenbauer, H., R. Schwenn, E. Marsch, B. Meyer, H. Miggenrieder, M.D. Montgomery, K.H. Muehlhaeuser, W. Pilipp, W. Voges, S.M. Zink (1977) A survey on initial results of the HELIOS plasma experiment, *Journal of Geophysics - Zeitschrift für Geophysik*, 42 (6) 561-580.

Rouillard, A.P., J.A. Davies, B. Lavraud, R.J. Forsyth, N.P. Savani, D. Bewsher, D.S. Brown, N.R. Sheeley, C.J. Davis, R.A. Harrison, M. Lockwood, S.R. Crothers, C.J. Eyles (2010a) Intermittent release of small-scale transients in the slow solar wind: I. Remote sensing observations, *J. Geophys. Res.,* **115**, A04103, doi:10.1029/2009JA014471

Rouillard, A.P., B. Lavraud, J.A. Davies, L.F. Burlaga, N.P. Savani, R.J. Forsyth, J.-A. Sauvaud, A. Opitz, M. Lockwood, J.G. Luhmann, K.D.C. Simunac, A.B. Galvin, C.J. Davis, R.A. Harrison (2010b) Intermittent release of small-scale transients in the slow solar wind: II. In-situ evidence, *J. Geophys. Res.*, **115**, A04104, doi:10.1029/2009JA014472.

Schwadron, N.A. (2002) An explanation for strongly underwound magnetic field in co-rotating rarefaction regions and its relationship to footpoint motion on the sun, *Geophys. Res. Lett.*, **29**(14), 1663, pp8.1-8.14, doi: 10.1029/2002gl015028

Sheeley, N. R., Y.-M. Wang, S.H. Hawley, G.E. Brueckner, K.P. Dere, R.A. Howard, M.J. Koomen, C.M. Korendyke, D.J. Michels, S.E. Paswaters (1997) Measurements of Flow Speeds in the Corona Between 2 and 30 $R_\odot$, *Astrophys. J.*, **484**, 472-478, doi: 10.1086/304338

Smith, C. W., and J. L. Phillips (1996) The role of CMEs and interplanetary shocks in IMF winding angle statistics, *AIP Conf. Proc.*, **382** (1), 502-505, doi: 10.1063/1.51438

Smith, C. W., and J. L. Phillips (1997), The role of coronal mass ejections and interplanetary shocks in interplanetary magnetic field statistics and solar magnetic flux ejection, *J. Geophys. Res.*, **102** (A1), 249-261, doi:10.1029/96JA02678.

Smith, C.W., J. L'Heureux, N.F. Ness, M.H. Acuna, L.F. Burlaga, and J. Scheifele (1998) The ACE magnetic fields experiment. In The Advanced Composition Explorer Mission (pp. 613-632). Springer, Dordrecht. doi: 10.1007/978-94-011-4762-0_21

Smith, E. J. (2011), What causes the flux excess in the heliospheric magnetic field?, *J. Geophys. Res.,* **116**, A12101, doi:10.1029/2011JA016521.

Suess, S. T., and E. J. Smith (1996), Latitudinal dependence of the radial IMF component coronal imprint, *Geophys. Res. Lett.*, **23**, 3267–3270, doi:10.1029/96GL02908.

Viall, N.M., and A. Vourlidas (2015) Periodic density structures and the origin of the slow solar wind, *Astrophys. J.,* **807**, 176 (13pp)

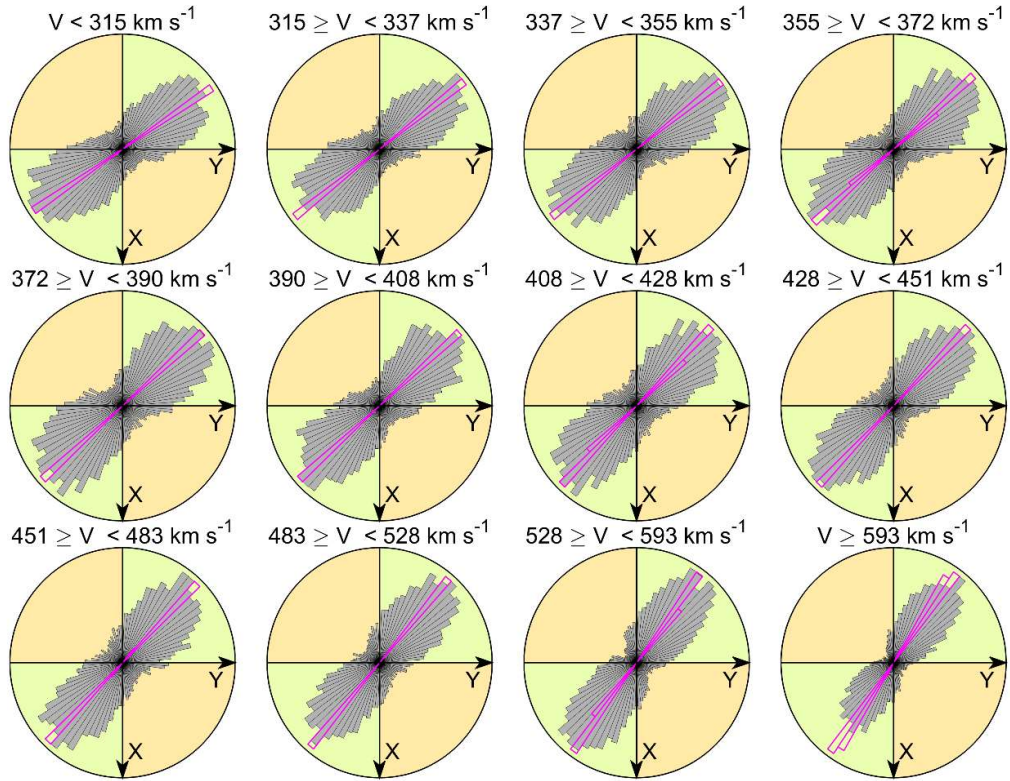

**Figure 1**. Occurrence distribution of near-Earth heliospheric field (HMF) orientation. The grey polar histograms show the distributions of the observed field garden hose angle θ (equation 1) in the GSE *X-Y* frame from 1-hour averages (τ = 1hr) of the HMF for the years 1996-2017 (inclusive, giving 161620 valid samples). The data are divided into 12 bins of equal sample numbers (13468 in each) between the percentiles of the distribution of the radial solar wind speeds, *V* which are 315, 337, 355, 372, 390, 408, 428, 451, 483, 528 and 593 km s$^{-1}$. The mauve histograms show the corresponding distributions of the orientation predicted from Parker spiral theory (θ$_p$, equation 4) for the observed hourly mean radial solar wind speed *V*. The rotation of the means and modes of the observed garden distribution towards the X axis with increasing *V* matches well that in the predicted field orientation. The ortho-gardenhose (OGH) and gardenhose (GH) sectors are shaded orange and green, respectively. The toward the Sun (T) OGH field is in the positive Y, positive X sector and the away from the Sun (A) OGH field is the negative Y, negative X sector and both are seen in all velocity ranges being most common for the lowest *V* and less common for the highest *V*.

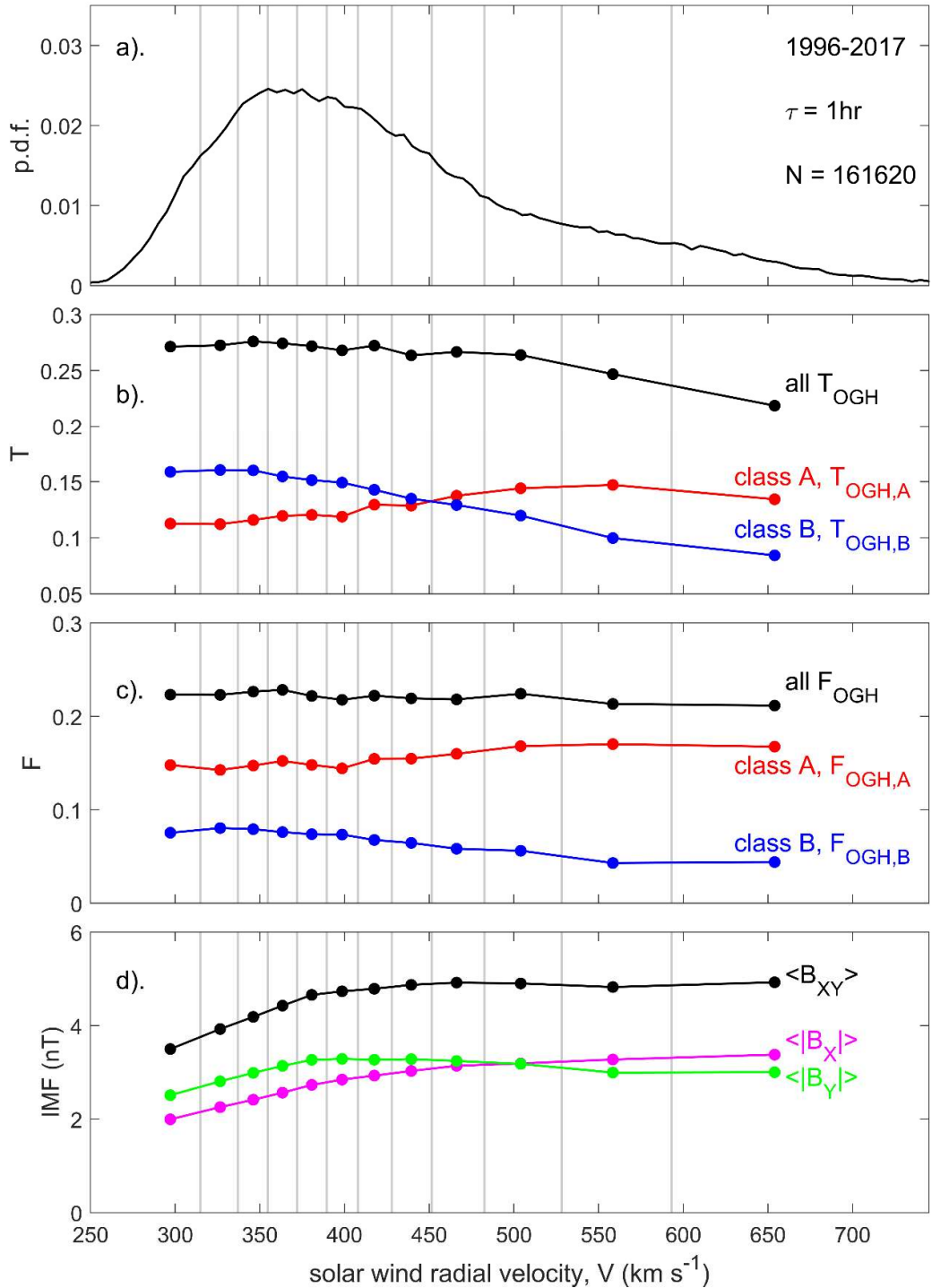

**Figure 2**. Effect of radial solar wind speed on averages of the near-Earth HMF. (a) the probability density function of the radial solar wind velocity, $V (= -V_X)$ for the $N = 161620$ hourly means ($\tau = 1$ hr) for 1965-2017 (inclusive). The vertical grey lines show the percentiles that give $N/12$ samples in each bin (as used in Figure 1). (b) The fraction of time $T$ in which the HMF is in an OGH orientation for each bin of $V$ (black points and line, $T_{OGH}$)

and for Class A and Class B rotations into that OGH orientation (red and blue points, $T_{OGH,A}$ and $T_{OGH,B}$, respectively). (c) The fraction $F_{OGH}$ of the total unsigned radial magnetic flux ($=\int|B_X|dt$) passing through unit width of the ecliptic plane at heliocentric distance $r = 1$AU that is in an OGH orientation (black line, $F_{OGH}$) and its two subdivisions of Class A and Class B OGH flux (red and blue lines, $F_{OGH,A}$ and $F_{OGH,B}$, respectively). (d) the average values for the $V$ bins of: (mauve) $|B_X|$, (green) $|B_Y|$, and (black) $B_{XY} = (B_X^2 + B_X^2)^{1/2}$ where $B_X$ and $B_Y$ are the $X$ and $Y$ components of the HMF in the GSE frame.

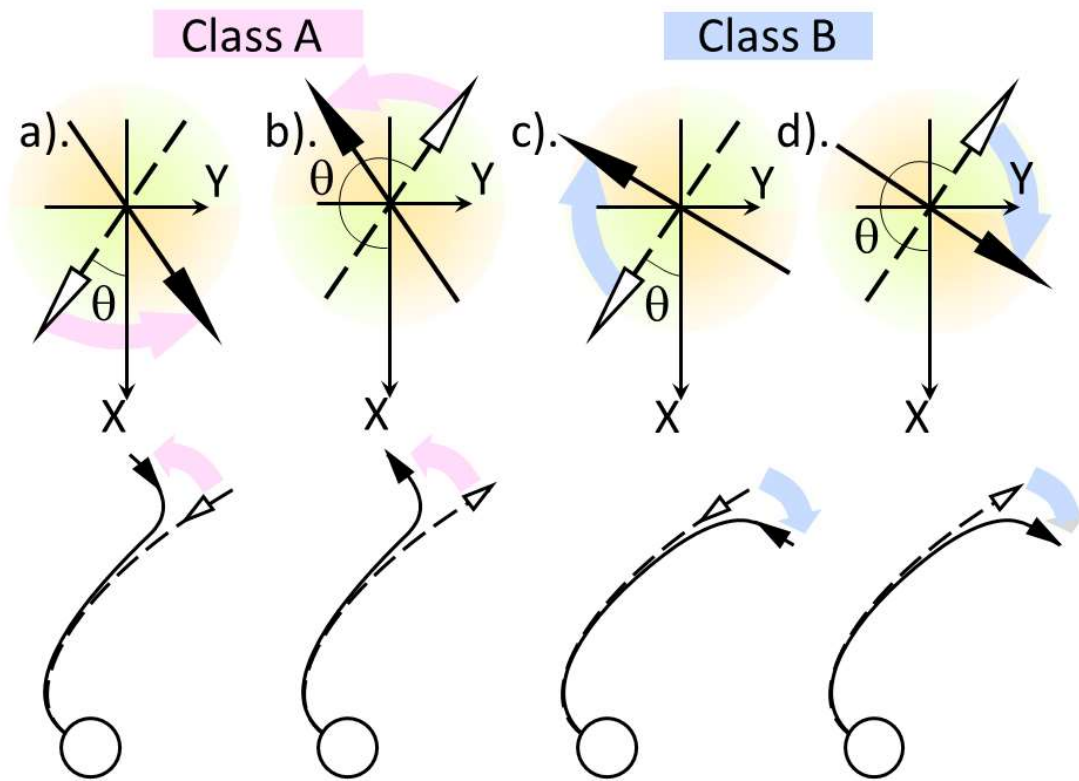

**Figure 3**. Schematic of θ sectors and HMF deviations from Parker spiral. The top panels show are local views of the near-Earth ecliptic plane from the north (i.e. looking in the –Z GSE direction): as in Figure 1, the OGH and GH quadrants are shaded orange and green, respectively. The lower panels are views of the inner heliosphere looking down from over the north pole of the Sun. In both cases, OGH field is shown by solid arrows with open and dashed arrows giving the ideal Parker spiral field direction. Parts (a) and (b) are termed Class A deflections of the local IMF vector from the Parker spiral with anti-clockwise rotation of the near-Earth field vector when viewed in the –Z direction (pink arrows); parts (c) and (b) are termed Class B deflections from the Parker spiral with clockwise rotation of the field vector when viewed in the –Z direction (pale blue arrows).  Note that in the lower panels the arrows show the displacement of the HMF field lines from the Parker spiral, not the sense of rotation.

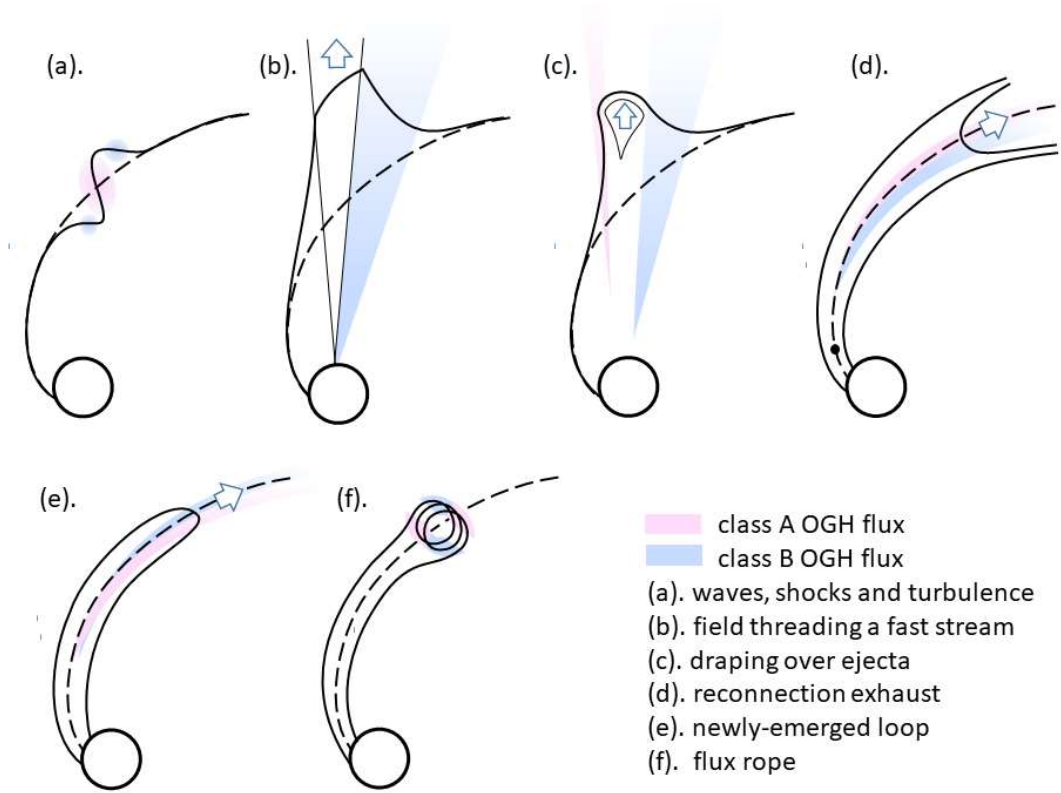

**Figure 4**. Schematics of some of the causes of OGH HMF orientations in near-Earth interplanetary space. The areas shaded pink and pale blue are snapshots of where Class A and Class B OGH flux (see Figure 3) exist at a given time and hence in cases (d) and (e), their radial extent depends on how long the reconnection and new loop emergence (respectively) persist. The dashed lines are an undisturbed parker spiral HMF field line orientation.

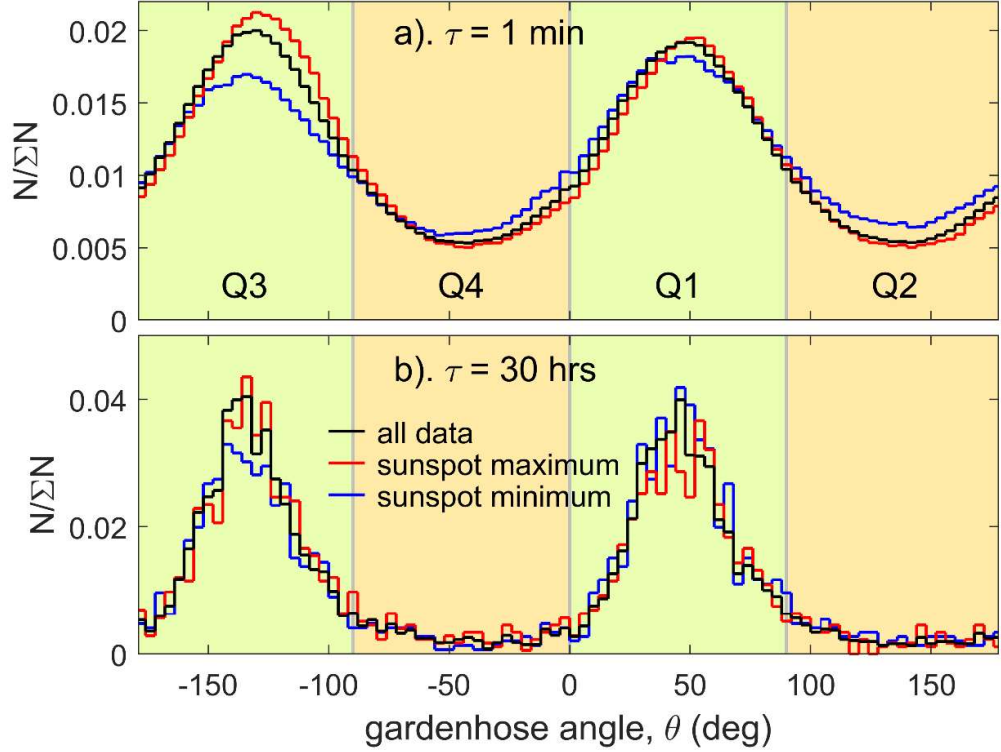

**Figure 5**. Occurrence distribution histograms as a function of observed gardenhose angle θ for the near-continuous IMF data available for 1996 to 2017 (inclusive): the black lines are for all data, red lines for the six years within 1 year of one of the two sunspot maxima and the blue lines are for the six years within one year of a sunspot minimum (which is here taken to include 2017). The number of valid samples in 4°-wide bins of θ are counted, $N$, and normalised to the sum for all 90 bins, $\Sigma N$. (a) is for an averaging timescale $\tau = 1$ min, and (b) is for $\tau = 30$ hrs. As in Figures 1 and 2, OGH and GH sectors are shaded orange and green, respectively. The four quadrants are labelled Q1 (T field in gardenhose orientation), Q2 (A field in ortho-gardenhose orientation), Q3 (A field in gardenhose orientation), and Q4 (T field in ortho-gardenhose orientation), where T and A are "Toward" and "Away" from the Sun, respectively ($B_X \geq 0$ and $B_X < 0$).

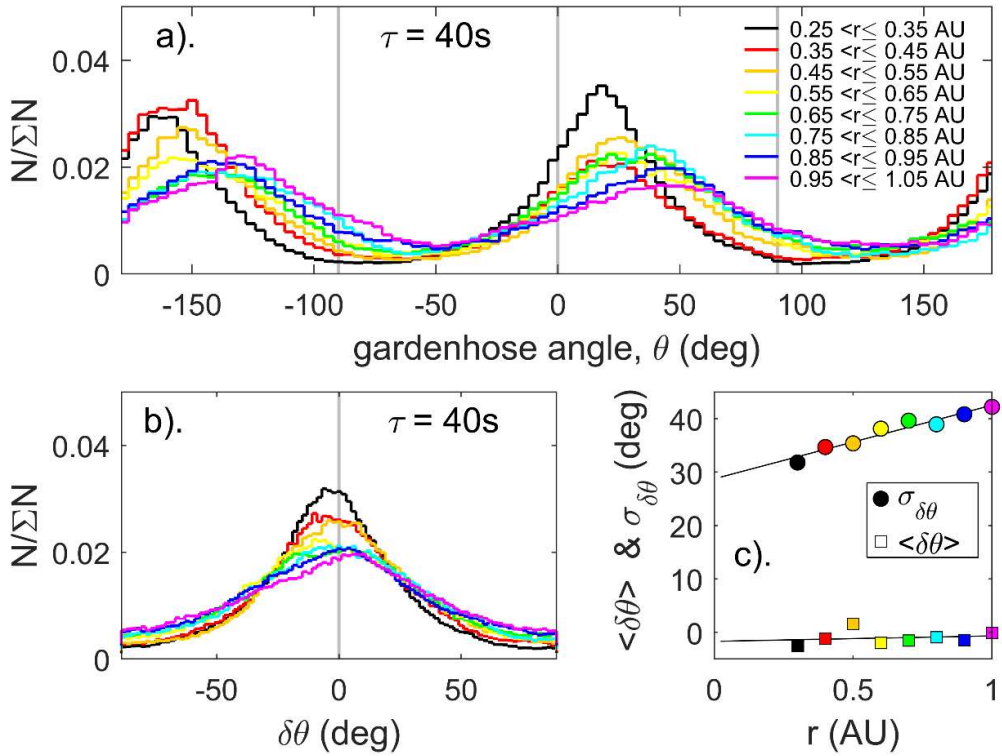

**Figure 6**. Helios-1 and Helios-2 data showing how the garden-hose angle distribution varies between 0.3 and 1 AU. Data are sorted into 8 non-overlapping bins of heliocentric distance, $r$, that are 0.1 AU wide and the centred on [0.3:0.1:1.0] AU. (a). Histograms of the normalised number of samples in bins of gardenhose angle $\theta$ that are 4º wide, $N/\Sigma N$, colour-coded by the range of $r$ using the key given. (b) The distributions of the deviations from the predicted Parker spiral angle $\delta\theta$ for the same ranges of $r$ and using the same colour scheme. (c) The variations with $r$ of the means and standard deviations of the $\delta\theta$ distributions shown in (b), respectively $\langle\delta\theta\rangle$ and $\sigma_{\delta\theta}$: the black lines are the best least-squares linear regression fits.

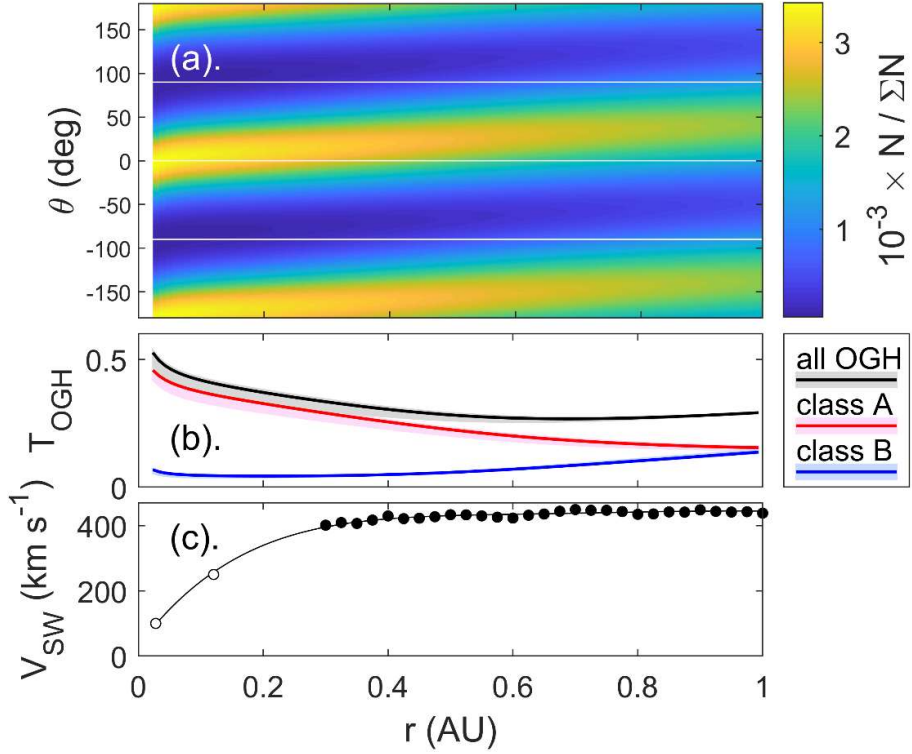

**Figure 7**. (a) Model of the evolution of the distributions of garden hose angle θ with heliocentric distance $r$ using Gaussian distributions of the deviations from the Parker spiral direction of mean $\langle\delta\theta\rangle$ and standard deviation $\sigma_{\delta\theta}$ given by the regression fits in Figure 6(c). The total toward and away distributions are assumed to be symmetrical and the linear variations in $\langle\delta\theta\rangle$ and $\sigma_{\delta\theta}$ seen at $r > 0.3$ are extrapolated to closer to the Sun. The Parker spiral direction at each $r$ is evaluated using the velocity variation $V(r)$ shown in part (c) which is a 5th–order polynomial fit to the observed average velocities in the Helios data (solid circles) and values at $r$ of $6R_\odot$ and $26R_\odot$ (open circles) derived by applying Fourier motion filters to SOHO/LASCO C3 movies observed from 1999 to 2010 by *Cho et al.* (2018). Panel (b) shows the predicted fraction of time that the field is in an OGH orientation, $T_{OGH}$: the black line is for all field, the red line for Class A OGH field and the blue line for Class B OGH field (see Figure 2). The areas shaded grey, pink and pale blue around the black, red and blue lines are the uncertainties introduced by the uncertainty in $\langle\delta\theta\rangle$.

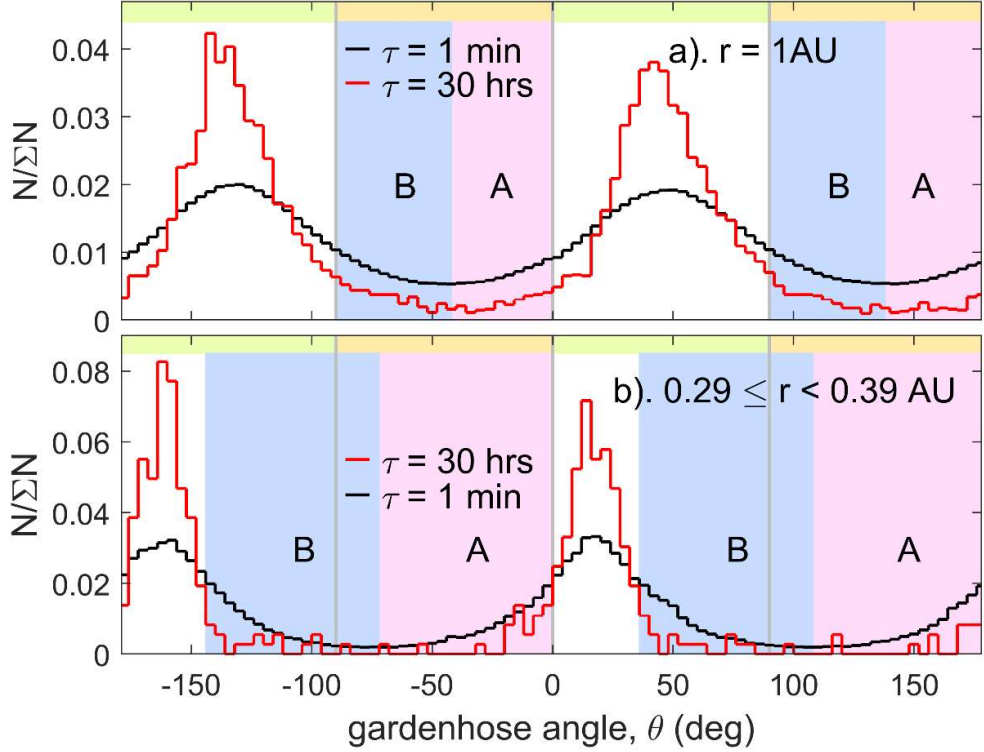

**Figure 8**. Comparison of probability density histograms ($N/\Sigma N$, where $N$ is counted in 4°-wide bins of HMF gardenhose angle $\theta$) at (a) $r = 1$AU (from the Omni dataset for 1996-2017, inclusive) and at (b) $r \leq 3.9$ AU (from the Helios 1 and 2 data for December 1974 to August 1981). In both cases $\theta$ has been determined using equation (@@) from 1-minute (in black) and 30-hour (in red) averages of HMF $B_X$ and $B_Y$ ($\tau = 1$ min and $\tau = 30$ hrs). The Helios data are taken from 40-s averages and linearly interpolated on to times one-minute apart: interpolation points that are separated from a valid 40-s mean by more than 40 seconds are treated as data gaps. For both datasets the 1-minute samples were averaged in 30 hour intervals and treated as valid means if the number of available 1-minute samples in each exceeded 1350, which is 75% of the maximum number of 1800. The orange and green bars at the top of each panel show ortho-gardenhose and garden hose orientations, respectively, and the pink and blue shading shows the definitions of Class A and Class B flux used in the remainder of this paper. For $r = 1$AU in (a), A and B are defined as two sub-classes of OGH flux, separated by the orthogonal to the average spiral direction. For the near-Sun field observed by Helios ($r < 0.29$ AU) the same definitions are not greatly informative as the smaller Parker spiral angle means that there would be almost no flux in Class B and a lot in Class A (as shown by Figure 7b): hence wider bands in $\theta$ are used, as defined in the text.

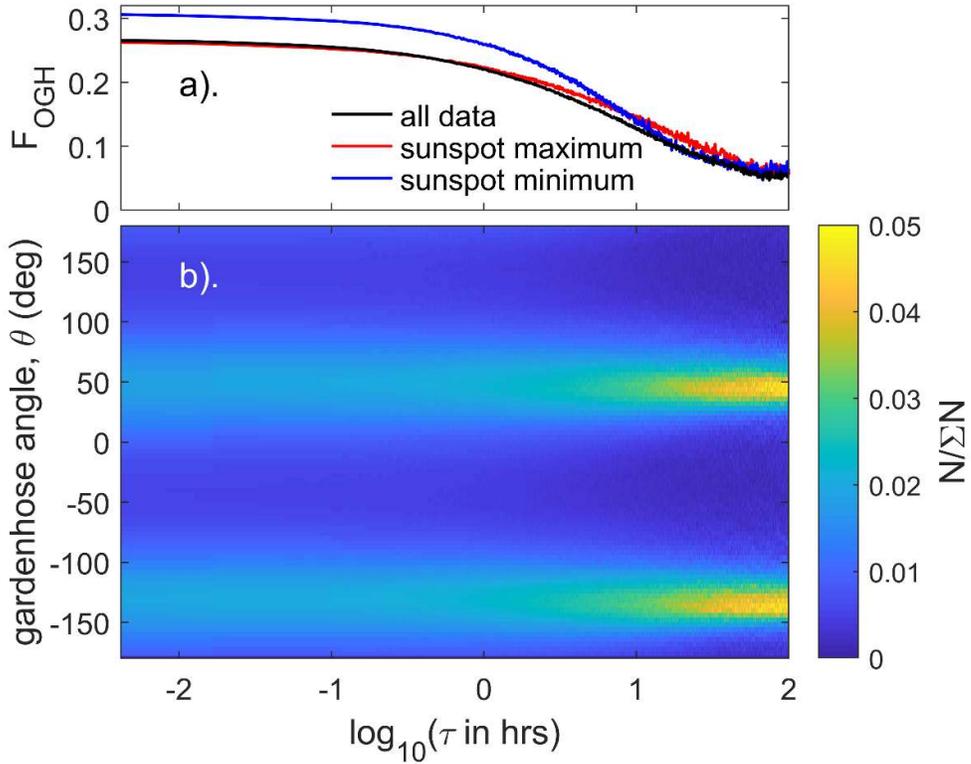

**Figure 9**. The effect of averaging on the distribution of gardenhose angle θ for $r$ = 1AU. For timescales $\tau \geq$ 1min ($\log_{10}(\tau$ in hrs) $\geq$ 1.77) the data used are 1-minute averages from the Omni dataset or 1996-2017. Plots are extended to $\tau$ below 1 minute using 15-second averages of data from the ACE and WIND spacecraft for the same interval. Part (b) shows $N/\Sigma N$ (where $N$ is counted in 4°-wide bins of HMF gardenhose angle θ) colour-coded as a function of $\log_{10}(\tau$ in hrs) and θ. Part (a) shows the fraction of the integrated total of the radial magnetic flux (toward or away) that is in a non-Parker-spiral Class A or Class B orientation (as defined by Figure 7a) $F$, as a function of $\log_{10}(\tau$ in hrs): the black line is for all years, the red line for three-year intervals around sunspot maxima and the blue line for three year intervals around sunspot minima.

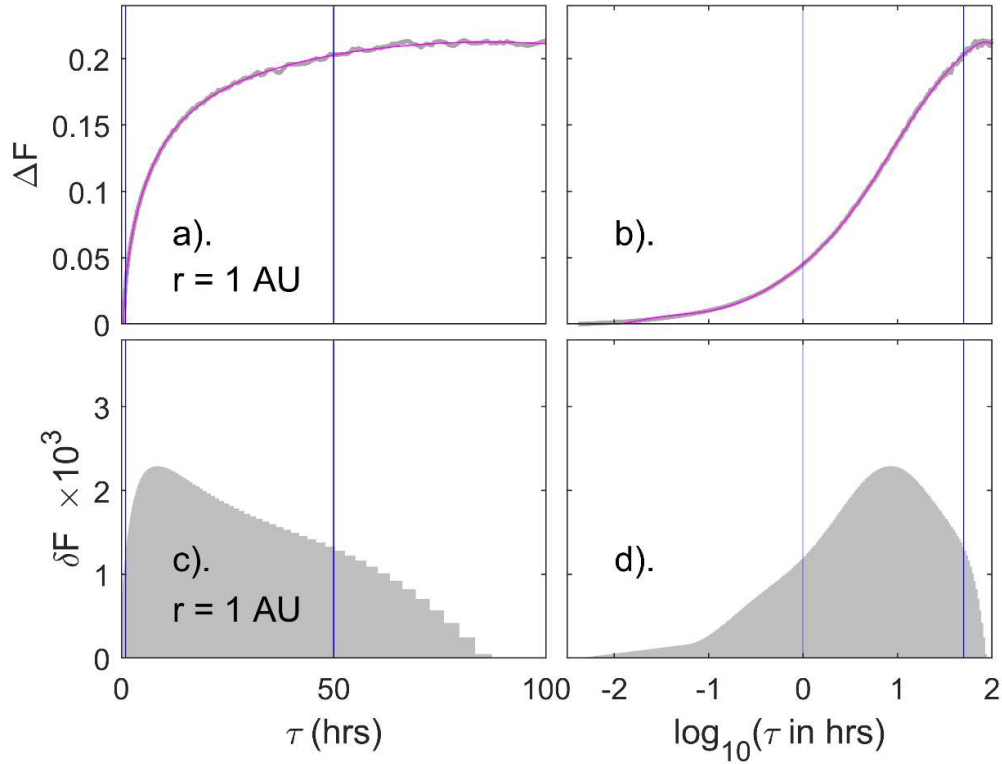

**Figure 10**. Analysis of the structure scale of all non-Parker-spiral (Class A and Class B flux combined) at $r$ = 1AU. The plots on the left hand side use an x axis that is linear in averaging timescale $\tau$, the right hand plots are the same, other than the x axis is logarithmic. Parts (a) and (b) are show $\Delta F$, the fraction of the total radial magnetic flux that is lost by cancellation by averaging over an interval $\tau$ (by definition zero for the fundamental resolution of the data which is $\tau$ = 1min.). The grey line is the observed value and the mauve line a 6$^{th}$ order polynomial fit to smooth out the small numerical noise associated with the start times of the averaging intervals relative to features in the data. (Note that the fit is generated using the $\log_{10}(\tau)$ variation shown in part b). Parts (c) and (d) show the flux lost in increasing the averaging time between $\tau_1$ and $\tau_2$, $\delta F$: in (c) the area shaded grey is made up of vertical bars of height $\delta F$ between $\tau_1$ and $\tau_2$ and in (d) the bars are again of height $\delta F$ but between $\log_{10}(\tau_1)$ and $\log_{10}(\tau_2)$. In this way, the total area shaded grey in both plots equals the total fraction of radial OGH flux in 1-minute data that is lost by averaging over intervals of duration $\tau$ = 100 hrs and the shapes of the grey areas allows us to identify what timescales $\tau$ contribute most to this loss, and hence gives the spectrum of scale sizes of the OGH flux. The linear plot (c) allows us to see structure at high $\tau$, the logarithmic plot (d) reveals structure at low $\tau$.

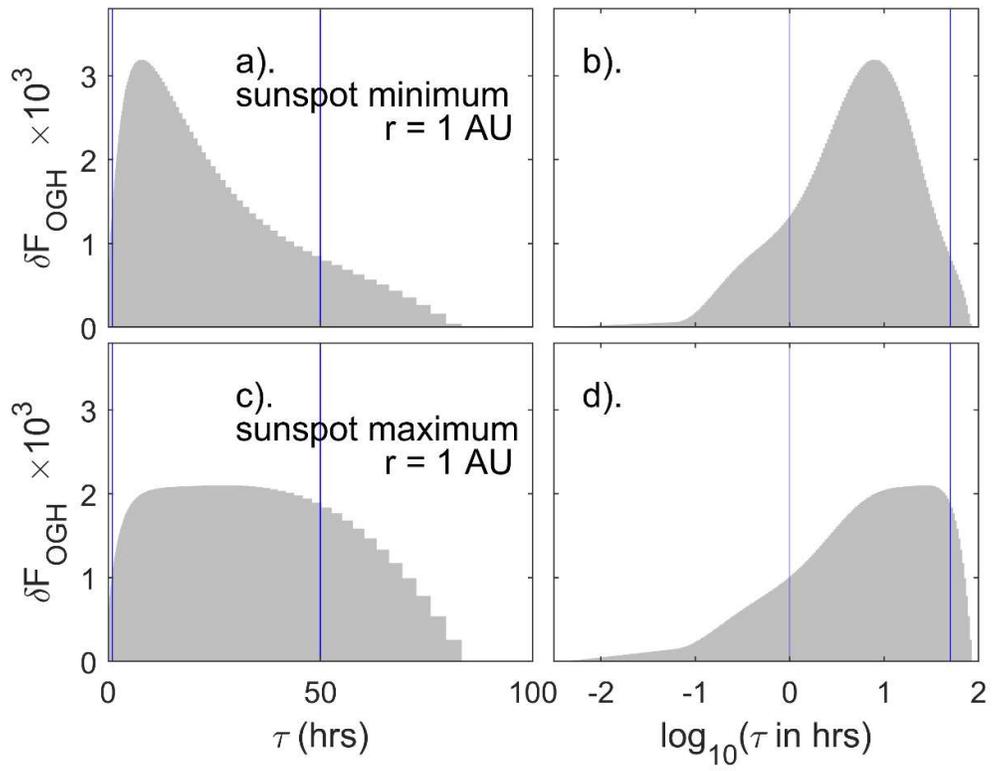

**Figure 11**. The same as Figures 10c and 10d for (top) data at $r = 1\text{AU}$ within one year of sunspot minimum and (bottom) one year of sunspot maximum.

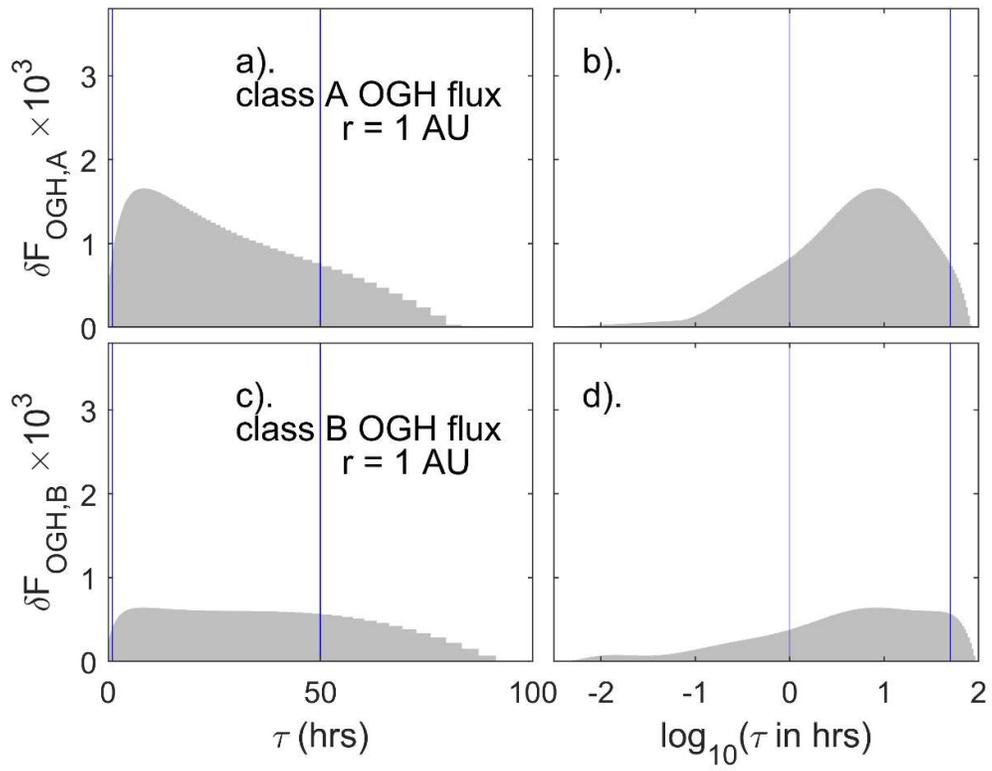

**Figure 12.** The same as Figures 10c and 10d for (top) Class A flux and (bottom) Class B flux at $r = 1\text{AU}$, as defined by the pink and pale blue areas, respectively, in figure 8a.

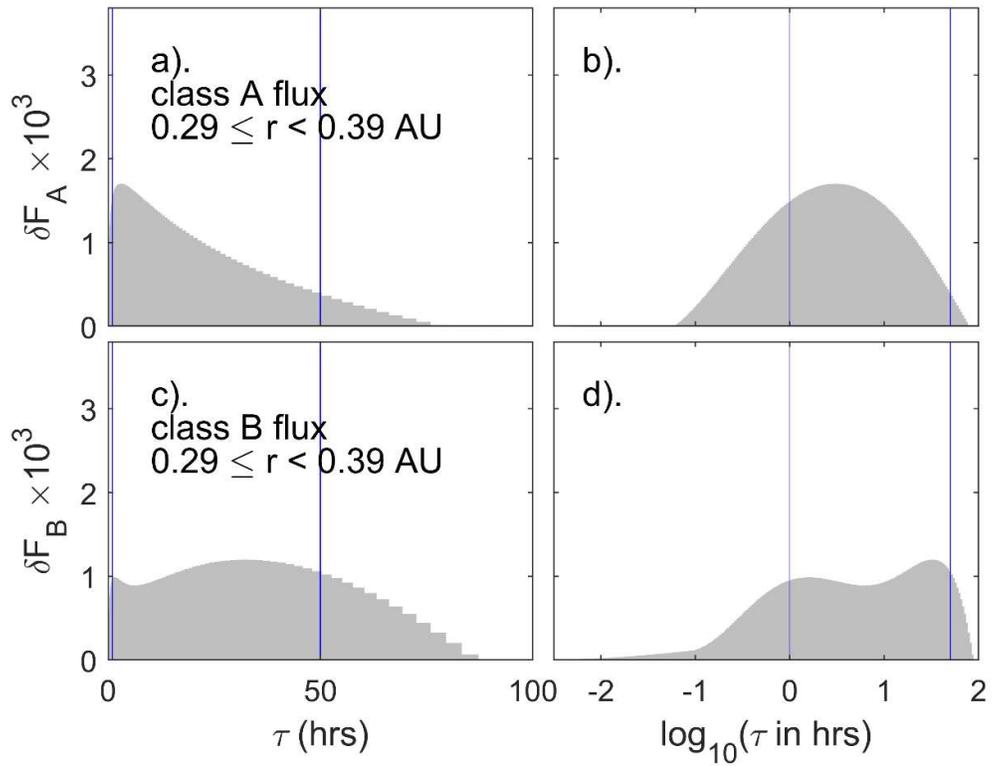

**Figure 13**. The same as Figures 10c and 10d for (top) Class A flux and (bottom) Class B flux observed by the Helios spacecraft at $r < 0.39$ AU, as defined by the pink and pale blue areas, respectively, in figure 8b.